\documentclass[
	pra,
	aps,
	reprint,
	amsmath,amssymb,
	a4paper,
	superscriptaddress,
	10pt,
    nofootinbib
]{revtex4-2}

\usepackage{times}      %
\usepackage{helvet}     %

\usepackage{physics}    %
\usepackage{upgreek}    %

\usepackage{graphicx}
\graphicspath{{fig/}}
\usepackage[dvipsnames]{xcolor}
\definecolor{rmpblue}{HTML}{2e3092}
\usepackage{booktabs}

\usepackage{enumitem}
\setlist[enumerate]{label=(\roman*)}
\setlist[itemize]{label=--}

\usepackage[
    colorlinks=true,
    citecolor=rmpblue,
    linkcolor=rmpblue,
    urlcolor=rmpblue,
    filecolor=magenta
]{hyperref}

\newcommand{\iu}{\mathrm{i}\mkern1mu}                 %
\newcommand{\eu}{\mathrm{e}\mkern1mu}                 %
\newcommand{\ort}[1]{\boldsymbol{\mathbf{\hat{#1}}}}  %
\newcommand{\um}{\ensuremath{\,\upmu\text{m}}}        %
\newcommand{\nm}{\ensuremath{\,\text{nm}}}            %

\newcommand{\affilANU}{Research School of Physics, Australian National University, Canberra ACT 2601, Australia}

\newcommand{\affilBombay}{Department of Physics, Indian Institute of Technology Bombay, Mumbai 400076, India}

\begin{document}

\title{Optical chirality of membrane metasurfaces with broken in-plane symmetry}

\author{Ivan Toftul}
\email{toftul.ivan@gmail.com}
\altaffiliation{equal contribution}
\affiliation{\affilANU}
\author{Brijesh Kumar}
\email{Brijesh.Kumar@iitb.ac.in}
\altaffiliation{equal contribution}
\affiliation{\affilANU}
\affiliation{\affilBombay}

%\author{Maxim Gorkunov}
%\affiliation{\relax}

\author{Yuri Kivshar}
\affiliation{\affilANU}

\date{\today}
\begin{abstract}
We study chiroptical properties of single-layer dielectric membrane metasurfaces with broken in-plane symmetry.
Although chiral optics was long tied to structures lacking any mirror symmetry, planar (2D) chiral structures that retain a horizontal mirror plane now attract growing attention for their strong chiral optical effects.
We explicitly show and explain how single-layer metasurfaces can achieve strong resonant chiral response.
Although the single-layer geometry forbids conventional co-polarized circular dichroism, strong resonant conversion circular dichroism appears to be possible in particular wavelength ranges determined by the spectra of photonic eigenmodes.
We explore its origin starting with a $C_4$ rotation-symmetric and in-plane mirror-symmetric membrane metasurface and applying to it various in-plane perturbations.
Simultaneous breaking of the in-plane mirror symmetry and lifting the rotation symmetry unlocks resonantly enhanced circular conversion dichroism.  We derive selection rules for this effect and trace its origin to eigenmode interference and intercoupling using chiral coupled-mode theory.
Material absorption weakens the conversion dichroism produced by modes of opposite up-down parity, but it is essential for that produced by modes of the same parity.
\end{abstract}

\maketitle

\section{Introduction}

Chirality, in its geometric sense, indicates non-identity of an object and its mirror image \cite{Kelvin1894Clarendon}. Optical chirality is a much more specific property which can be precisely quantified by the dissymmetry of interactions with circularly polarized light.
Historically, chiral optical phenomena are largely attributed to geometric chirality~\cite{fresnel_considerations_1824,Barron2012Chirality}.
Thus a differential optical response of chiral enantiomer solutions to left and right circularly polarized (LCP and RCP) light is generally referred to as \textit{circular dichroism} (CD) manifesting itself through dissymmetry of absorption, reflection and transmission of layers or interfaces~\cite{polavarapu2018chiral}, as well as of
scattering cross sections of compact single objects \cite{bustamante_circular_1983}.

\begin{figure}
	\centering
	\includegraphics[width=\linewidth]{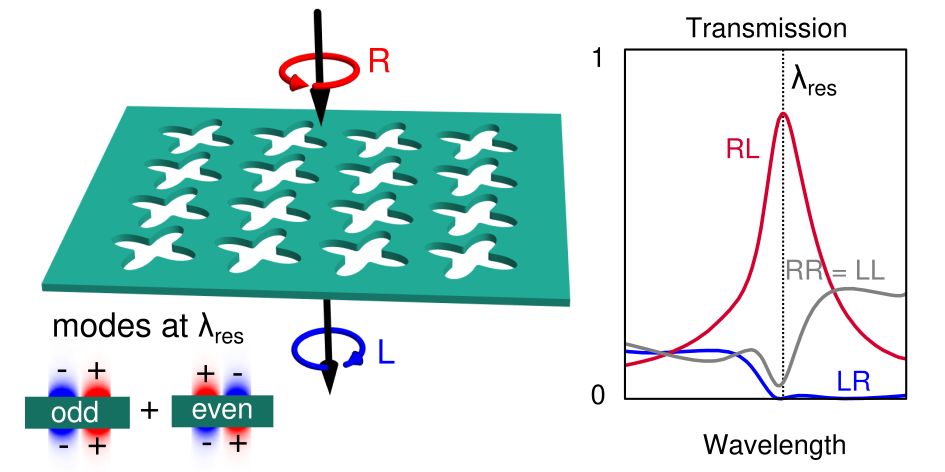}
	\caption{Concept of circular conversion dichroism.}
	\label{fig:toc}
\end{figure}

Artificial chiral electromagnetic materials have evolved from upscale models explaining the origin of natural optical chirality \cite{bose_rotation_1898,lindman_uber_1920}. Nowadays, as the demand for sources, detectors, filters, and sensors of chiral light is growing, compact chiral optical elements are extensively studied. Here the concept of chiral metasurfaces -- two-dimensional subwavelength arrays of chiral meta-atoms -- attracts significant attention. Such metasurfaces built of meta-atoms of high refractive index transparent materials demonstrate strongly chiral high-quality factor (high-Q) resonances underpinned by specific photonic eigenmodes~\cite{Gorkunov2020PRL,zhang_chiral_2022,chen_observation_2023}.

While the shapes of meta-atoms can be relatively simple, the in-plane mirror symmetry can be broken by multiple means: from breaking symmetry on the meta-atom level~\cite{Koshelev2023ACSPhot}, rotation of simple shapes relative to the lattice arrangement~\cite{Sinev2025NC,Gryb2023NL}, monoclinic lattice arrangement~\cite{Toftul2024PRL}, as well as using anisotropic materials by achieving a mismatch between the optical axis and the lattice arrangement~\cite{Wang2025PRL,An2026AFM,Zhao2022PhotRes}.
The out-of-plane (vertical)  mirror symmetry is broken by different heights~\cite{Gorkunov2020PRL,Kuhner2023LSA,Gorkunov2021AOM,Heimig2026SA}, by a lateral tilt~\cite{Zhang2022S,Chen2023N}, non-ideal fabrication~\cite{Glebov2026arXiv}, or by multilayer design~\cite{Gromyko2025PRL,Gromyko2024NC}. Although in some cases, this symmetry breaking can be more subtly produced by a substrate~\cite{Gorkunov2025AOM,Toftul2024PRL} or adjacent layers~\cite{Kumar2025ACSPhot,kumar_intrinsic_2026}, the resulting chiral optical phenomena are all determined by the fully broken mirror symmetry. The main effect in transmission is then designated by the co-polarized circular dichroism:
\begin{equation}
    \mathrm{CD}_{\mathrm{co}} = \frac{T_{\text{RR}} - T_{\text{LL}}}{T_{\text{RR}} + T_{\text{LL}}}.
\label{eq:CDco}
\end{equation}
The energy transmission coefficients $T$ here are defined in the basis of circular polarizations with the second index denoting the polarization of the incident wave and the first one -- the polarization of the outgoing transmitted wave. Note that even for a geometrically chiral metasurface $\mathrm{CD}_{\mathrm{co}}$~\eqref{eq:CDco} can vanish if it possesses a rotation symmetry of the order 3 or higher and its materials are reciprocal and not sufficiently strongly light absorbing~\cite{Gorkunov2024,Kumar2025ACSPhot}.

At the same time, it is regularly noted that certain chiral optical phenomena can be present in flat metasurfaces retaining their out-of-plane mirror symmetry~\cite{Hu2025APR,Semnani2020, shi_planar_2022, Voronin2022}.
Being geometrically achiral, such structures are incapable of conventional $\mathrm{CD}_{\mathrm{co}}$, but they can perform as resonant chiral mirrors reflecting waves of one helicity and transmitting the others while fully flipping their helicity. This type of chiral optical dissymmetry is known as \textit{circular conversion dichroism}~\cite{Fedotov2006PRL}
defined similarly to Eq.~\eqref{eq:CDco} as:
\begin{equation}
    \mathrm{CD}_{\mathrm{cross}}= \frac{T_{\text{RL}} - T_{\text{LR}}}{T_{\text{RL}} + T_{\text{LR}}}.
\label{eq:CDcross}
\end{equation}
For applications involving chiral light manipulations, the transmittance difference:
\begin{equation}
    \Delta T_{\mathrm{cross}} = T_{\text{RL}} - T_{\text{LR}},
\label{eq:dTcross}
\end{equation}
is more relevant.
In this work we keep vertical symmetry on purpose. This isolates the effect of in-plane symmetry breaking alone. In less symmetric structures that also lack the vertical mirror symmetry, $\mathrm{CD}_{\mathrm{co}}$ and $\mathrm{CD}_{\mathrm{cross}}$ generally coexist~\cite{Gryb2023NL}.

Note that although circular polarization conversion is generally performed also by birefringent objects (e.g.\ half-wave plates), for them always $\Delta T_{\mathrm{cross}} = 0$. Early plasmonic metamaterial-based designs have demonstrated the possibility of imbalance between the conversion directions~\cite{Fedotov2006PRL,Fedotov2007NL,Plum2009APL}. Dielectric-based designs less suffering from absorption losses  emerged almost a decade later~\cite{Wu2014NC,Hu2017SR,Semnani2020,Voronin2022, Gryb2023NL}.

\begin{figure*}
	\centering
	\includegraphics[width=\linewidth]{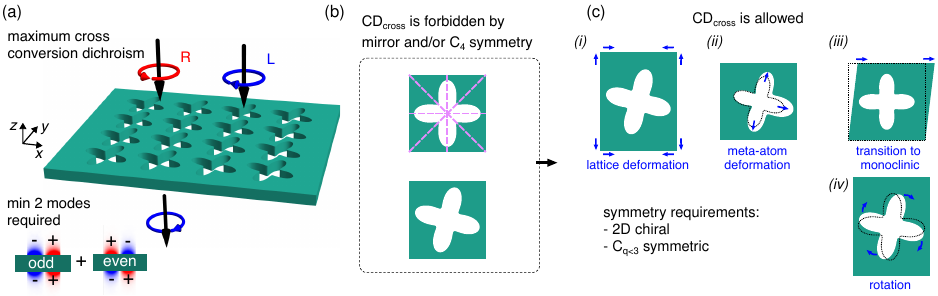}
	\caption{
		\textbf{Cross conversion induced by in-plane symmetry breaking.}
		{(a)} A single-layer dielectric membrane metasurface composed of a rectangular lattice of rotated $C_{4}$-symmetric holes supports resonantly enhanced polarization conversion.
        Colors in the left bottom insets represent out-of-plane electric field (blue negative, red positive).
		Two conditions have to be satisfied to enable the cross conversion: \textit{(1)} break all in-plane mirror symmetries and \textit{(2)} have $C_{q<3}$ rotationally symmetric structure.
		{(b--c)} Examples of symmetry-reduced unit cells are shown. Panels (i)--(iv) illustrate how lattice or meta-atom deformations transform the structure into a two-dimensionally chiral ($2$D-chiral, $C_{q}$-symmetric) system where CD$_{\mathrm{cross}}$ becomes allowed.
	}
	\label{fig:concept}
\end{figure*}

Despite the practical relevance and a number of works studying particular designs, a systematic theoretical analysis of the resonant conversion circular dichroism in dielectric metasurfaces has been largely overlooked, and this is the focus of the present work.
From the symmetry point of view, a nonzero $\mathrm{CD}_{\mathrm{cross}}$ \eqref{eq:CDcross} at normal incidence requires two conditions to be simultaneously satisfied  (Fig.~\ref{fig:concept}):
\begin{enumerate}
    \item \label{cond:rotation} the combined rotation-symmetry order of the meta-atom and lattice must be lower than 3;
    \item \label{cond:mirror} the structure must lack in-plane mirror symmetry.
\end{enumerate}
The first condition follows from the rotational invariance of the metasurface S-matrix, which forces all cross-polarized transmission coefficients to vanish for the rotation symmetry axis $C_{{q} \geq 3}$ (consequence of Hermann's theorem~\cite{Hermann1934}). The second condition ensures that left- and right-circular channels are not related by a mirror transformation. We discuss both in more detail in Section~\ref{sec:selectionrules}.

To systematically analyze the origin of $\mathrm{CD}_{\mathrm{cross}}$, we start with single-layer membrane metasurfaces for which it is forbidden by at least one of the two conditions: the $C_4$ rotation symmetry and an in-plane mirror-symmetry.
Applying various in-plane perturbations we simultaneously break the in-plane mirror symmetry and the rotation symmetry and unlock resonantly enhanced $\mathrm{CD}_{\mathrm{cross}}$.
As the optical properties of dielectric metasurfaces are largely determined by their resonances underpinned by excitation and emission of certain photonic eigenmodes, we analyze the underlying modal band structure and show how the observable $\mathrm{CD}_{\mathrm{cross}}$
can be reduced to eigenmode interference and intercoupling described in  terms of a coupled-mode theory (CMT). As we demonstrate in Section~\ref{sec:chiralCMT}, one can impose a third more specific condition:
\begin{enumerate}[start=3]
    \item \label{cond:overlap} for polarization-isotropic non-resonant background, at least two resonances must spectrally overlap.
\end{enumerate}
Unlike the first two, it emerges from the structure of the resonant coupling: for a membrane with polarization-isotropic background scattering, no spectrally isolated resonance can produce cross conversion at normal incidence, hence the cooperation of at least two overlapping modes is always required.
To the best of our knowledge, this requirement has not been explicitly identified in prior works, although it has in fact always been implicitly satisfied~\cite{Plum2009APL,Fedotov2006PRL}.
We also show that, depending on whether the interacting modes have opposite or equal parity, two distinct physical mechanisms of cross conversion arise. As test platforms we use perforated membranes, and we progressively reduce their symmetry by deforming the lattice. We observe the emergence of nonzero cross conversion (Fig.~\ref{fig:concept}).

We note that the above constraints significantly differ if light is obliquely incident on a metasurface, and in particular, a spectral overlap of multiple eigenmodes is not required~\cite{Fan2025OE,Liu2019PRL,Plum2008APL,Plum2009PRL}. This is beyond the scope of the present work as well as cases involving birefringent and/or optically active materials which determine trivial anisotropic nonresonant polarization conversion.

\section{Symmetry selection rules for conversion circular dichroism}
\label{sec:selectionrules}

In this section we address conditions~\ref{cond:rotation} and~\ref{cond:mirror}.
Throughout, we work with the monochromatic fields with oscillation frequency $\omega$ and adopt the $\eu^{-\iu\omega t}$ time convention, so that a decaying resonance has a \text{negative} imaginary part.
In the absence of diffraction, the transmission-reflection problem is described compactly by the scattering matrix in the circular polarization basis
\begin{equation}
    \ort{e}_{\pm} = \frac{\ort{x} \pm \iu\, \ort{y}}{\sqrt{2}}.
    \label{eq:circbasis}
\end{equation}
Here $\ort{x}$ and $\ort{y}$ are the unit orts along $x$- and $y$-axis. Mapping between the $\pm$ components and the right/left circular polarization (RCP/LCP) depends on the propagation direction. For a wave travelling along $+z$, $\ort{e}_{-}$ ($\ort{e}_{+}$) corresponds to RCP (LCP), whereas for $-z$ propagation the assignment is reversed.
We denote incident and outgoing channel amplitudes by $\vb{a} = (a_{\text{R}}, a_{\text{L}}, a_{\text{R}}^{\prime}, a_{\text{L}}^{\prime})^{\mathsf{T}} = (a_{-}, a_{+}, a_{+}^{\prime}, a_{-}^{\prime})^{\mathsf{T}}$ and $\vb{b} = (b_{\text{R}}, b_{\text{L}}, b_{\text{R}}^{\prime}, b_{\text{L}}^{\prime})^{\mathsf{T}} = (b_{+}, b_{-}, b_{-}^{\prime}, b_{+}^{\prime})^{\mathsf{T}}$, with primes marking the back ($z<0$) side of the metasurface.
The scattering matrix $\vb{b} = \vb{S}\cdot \vb{a}$ then reads~\cite{Gorkunov2020PRL,Gorkunov2024,Koshelev2024JOPT}
\begin{equation}
    \vb{S} =
    \begin{pmatrix}
        r_{\text{RR}} & r_{\text{RL}} & t^\prime_{\text{RR}} & t^\prime_{\text{RL}} \\
        r_{\text{LR}} & r_{\text{LL}} & t^\prime_{\text{LR}} & t^\prime_{\text{LL}} \\
        t_{\text{RR}} & t_{\text{RL}} & r^\prime_{\text{RR}} & r^\prime_{\text{RL}} \\
        t_{\text{LR}} & t_{\text{LL}} & r^\prime_{\text{LR}} & r^\prime_{\text{LL}}
    \end{pmatrix},
    \label{eq:Smatrix}
\end{equation}
where the second index denotes the polarization of the incident wave and the first index the polarization of the outgoing wave.
The corresponding power coefficients are $T_{ij}=|t_{ij}|^2$ and $R_{ij}=|r_{ij}|^2$, with $i,j\in\{\mathrm{R},\mathrm{L}\}$.
Lorentz reciprocity requires~\cite{Koshelev2024JOPT}
\begin{equation}
\begin{split}
    t_{\text{RR}} &= t_{\text{RR}}^{\prime}, \quad
    t_{\text{LL}} = t_{\text{LL}}^{\prime}, \quad
    t_{\text{RL}} = t_{\text{LR}}^{\prime}, \\
    t_{\text{RL}}^{\prime} &= t_{\text{LR}}, \quad
    r_{\text{RL}} = r_{\text{LR}}, \quad
    r_{\text{RL}}^{\prime} = r_{\text{LR}}^{\prime},
\end{split}
    \label{eq:reciprocity}
\end{equation}
while the up-down mirror symmetry $\hat{\sigma}_z$ of a single-layer membrane additionally forces $t_{\text{RR}} = t_{\text{LL}}$. Together, these reduce the S-matrix to six independent complex parameters.
The equality $t_{\text{RR}} = t_{\text{LL}}$ forces $\mathrm{CD}_{\mathrm{co}} = 0$~\eqref{eq:CDco}, while $t_{\text{RL}}$ and $t_{\text{LR}}$ remain independent, so any chiral response of a membrane must originate from the cross-polarized channel~\eqref{eq:CDcross}.

For a two-dimensional periodic structure, the allowed rotational symmetry orders are $q = 1, 2, 3, 4, 6$. Rotational order $q \geq 3$ forces cross-polarized transmission to zero, $t_{\text{LR}} = t_{\text{RL}} = 0$, and likewise forbids helicity-preserving reflection, $r_{\text{RR}} = r_{\text{LL}} = 0$~\cite{Koshelev2024JOPT,Gorkunov2024}. Any in-plane mirror symmetry independently suppresses cross-conversion as well. Hence, to support $\mathrm{CD}_{\mathrm{cross}}$, a membrane metasurface must belong to the $C_1$ or $C_2$ point group with all in-plane mirrors broken.

Importantly, these conditions are necessary but not sufficient. Even in a $C_1$- or $C_2$-symmetric membrane with all mirror symmetries broken, a single spectrally isolated resonance cannot produce $\mathrm{CD}_{\mathrm{cross}}$.
The up-down mirror symmetry of a flat membrane constrains the mode coupling coefficients in a way that for a spectrally isolated resonance cross-conversion dichroism is not possible. The interplay of at least two resonant modes is always required, as we show in the next section.

\section{Chiral coupled-mode theory for flat membranes}
\label{sec:chiralCMT}

To understand why condition~\ref{cond:overlap} is necessary and to identify the physical mechanisms of cross conversion, we employ temporal coupled-mode theory (CMT).

We consider periodic systems without diffraction. Hence, there are only 4 orthogonal outgoing channels (ports). We choose the basis of circular polarizations and describe each mode's coupling to the radiative channels by a vector of coupling parameters $\vb{m}_n = (m_{n\mathrm{R}},\, m_{n\mathrm{L}},\, m'_{n\mathrm{R}},\, m'_{n\mathrm{L}})^{\mathsf{T}}$, where the indexes $R,L$ denote the particular circular polarization and the primes distinguish the back ($z<0$) metasurface side~\cite{Toftul2024PRL,Gorkunov2020PRL,Gorkunov2025AOM,Kumar2025ACSPhot}.
The full $4 \times N$ coupling matrix of all considered states is built as $\vb{M} = (\vb{m}_1, \dots, \vb{m}_N)$.
One can explicitly express the coupling parameters by overlap integrals of the modal and port field distributions when the modes are the solutions of homogeneous Maxwell's equations used in resonant state expansions (RSEs)~\cite{weiss_how_2018,Kumar2025ACSPhot,Toftul2024PRL}.
As we discuss below, the modes entering CMT are not always identical with those in RSE, see also Refs.~\cite{Wu2024OE,Zhang2020arXiv}. However, as we show in Section~\ref{app:CMTC} of the Supporting Information, the implications of symmetry, electromagnetic reciprocity and reversibility substantially reduce the number of unknown phenomenological constants determining the coupling matrix elements.

The temporal evolution of the slow-varying mode amplitudes $\boldsymbol{\psi} = (\psi_1, \dots, \psi_N)^{\mathsf{T}}$, driven by monochromatic excitation at frequency $\omega$, is governed by~\cite{Suh2004,Gorkunov2024}
\begin{equation}
    \frac{\dd\boldsymbol{\psi}}{\dd t} = \iu( \omega\mathbf{I} - \mathbf{\Omega} + \iu \mathbf{\Gamma} )\,\boldsymbol{\psi} + \vb{M}^{\mathsf{T}}\,\vb{a},
    \label{eq:CMT1main}
\end{equation}
where $\vb{a}$ is the vector of incident wave amplitudes, $\mathbf{I}$ is an identity matrix, $\mathbf{\Omega}$ containing the resonance frequencies on its diagonal, and $\mathbf{\Gamma}$ encoding radiative decay on its diagonal and far-field (port-non-orthogonal) coupling via its off-diagonal elements~\cite{Suh2004}:
\begin{equation}
    \mathbf{\Omega} = \begin{pmatrix}
        \omega_{1}  & 0 & \dots & 0\\
        0 & \omega_{2}  & \dots & 0\\
        \vdots  & \vdots & \ddots & \vdots \\
        0 & 0 & \dots & \omega_{N}
    \end{pmatrix}, \quad
    \mathbf{\Gamma} = \begin{pmatrix}
       \gamma_1 &   \gamma_{12} & \dots &  \gamma_{1N}\\
         \gamma_{21} & \gamma_2 & \dots &  \gamma_{2N}\\
        \vdots  & \vdots & \ddots & \vdots \\
         \gamma_{N1} &  \gamma_{N2} & \dots &  \gamma_N
    \end{pmatrix}.
    \label{eq:OmegaGamma}
\end{equation}
An off-diagonal element $\gamma_{n\ell}$ in \eqref{eq:OmegaGamma} can be nonzero when modes $n$ and $\ell$ radiate into a shared combination of ports. This becomes unavoidable when the number of modes sharing some ports exceeds the number of the ports~\cite{Suh2004}, but it also happens for two modes, as we show below.
We term two $n$ and $\ell$ CMT modes \textit{port-non-orthogonal} when their $\gamma_{n\ell}\neq 0$ and \textit{port-orthogonal} when $\gamma_{n\ell}=0$.
A non-diagonal $\boldsymbol{\Gamma}$ implies, according to Eq.~\eqref{eq:CMT1main}, that the CMT modes mix even without external perturbation. %
Therefore, they are distinct from the true eigenmodes found by numerical solvers (quasinormal modes), which correspond to the poles of the scattering matrix or exact solutions of the eigenvalue problem.

\begin{table*}
  \centering
  \caption{\textbf{Bare and dressed modes summary.} The CMT coordinates of
    Eq.~\eqref{eq:CMT1main} and the S-matrix poles of Eq.~\eqref{eq:SmatCMTmain}
    are two different objects in principle. Bare modes are the natural variables of
    CMT theory, while dressed modes are what numerical eigensolvers and experiments
    actually see. The two coincide only when $\vb{\Gamma}$ is diagonal. The last
    row assumes no losses; with losses, same-parity modes also contribute
    (Supporting Information, Section~\ref{app:SmatPortNonOrthog}).}
  \label{tab:bare_vs_dressed}
  \begin{tabular}{p{0.3\linewidth}p{0.34\linewidth}p{0.34\linewidth}}
    \toprule
                           & \multicolumn{1}{c}{\textbf{Bare modes}}
                           & \multicolumn{1}{c}{\textbf{Dressed (quasinormal) modes}} \\
    \midrule
    Definition             & Enter CMT, Eq.~\eqref{eq:CMT1main}
                           & S-matrix poles, Eq.~\eqref{eq:SmatCMTmain} \\[3pt]
    How to access          & Fit CMT to data
                           & Numerical eigensolver \\[3pt]
    Coupling matrix        & $\vb{M}$ obeys \eqref{eq:timereversalconstraint},\,\eqref{eq:energyconstraint}
                           & $\tilde{\vb{M}}$ may not obey \eqref{eq:timereversalconstraint},\,\eqref{eq:energyconstraint} \\[3pt]
    Far-field polarization & Linear, Eq.~\eqref{eq:mRmLequal}
                           & Elliptical \\
    \midrule
    Mirror parity          & \multicolumn{2}{c}{Well-defined} \\[3pt]
    Coincide when          & \multicolumn{2}{c}{Port-orthogonal limit ($\gamma_{n\ell}=0$, $n\neq \ell$)} \\[3pt]
    $\Delta T_{\mathrm{cross}}$ origin
                           & \multicolumn{2}{c}{Interference of opposite-parity modes} \\
    \bottomrule
  \end{tabular}
\end{table*}

The outgoing wave amplitudes are determined by
\begin{equation}\label{eq:CMT2main}
	\vb{b} = \vb{M}\,\boldsymbol{\psi} + \vb{C}\,\vb{a},
\end{equation}
where the first term accounts for re-radiation by the resonant modes and the second describes non-resonant background scattering through the matrix $\vb{C}$.
The same coupling matrix $\vb{M}$ appears in both equations, as required by Lorentz reciprocity~\cite{Suh2004}. In the absence of light energy dissipation, one can further require the energy conservation and the time-reversal symmetry of CMT equations. This yields additional constraints:
\begin{align}
    \vb{C} \vb{M}^{*} = - \vb{M}, \label{eq:timereversalconstraint} \\
    2 \vb{\Gamma} = \vb{M}^{\dagger} \vb{M}. \label{eq:energyconstraint}
\end{align}
Identity~\eqref{eq:energyconstraint} states that all decay rates are determined by the energy leaking out through the radiative ports. Therefore, $\vb{\Gamma}$ here is, by construction, the radiative decay matrix, $\vb{\Gamma}\equiv\vb{\Gamma}_{\mathrm{rad}}$.
We hold to this definition throughout this work.
The constraints~\eqref{eq:timereversalconstraint} and~\eqref{eq:energyconstraint} hold true only for $\vb{\Gamma}_{\mathrm{rad}}$.
Absorption may affect all CMT parameters including complex eigenfrequencies and coupling parameters. However, for weak absorption, the corresponding corrections to $\vb{\Omega}$ and $\vb{M}$ are also weak and can be neglected in the first approximation. This is different for the matrix $\vb{\Gamma}$ which, for high-quality factor modes, consists of small decay rates.  Accordingly, we assume that the main absorption effects are reduced to the changes of the matrix  $\vb{\Gamma}$. The role of absorption is discussed in more detail below in Sec.~\ref{sec:losses}.

For monochromatic steady-state excitation ($\dd\boldsymbol{\psi}/\dd t = 0$), eliminating $\boldsymbol{\psi}$ from Eqs.~\eqref{eq:CMT1main} and~\eqref{eq:CMT2main} yields the scattering matrix (see Supporting Information, Section~\ref{app:CMTC}):
\begin{equation}
	\vb{S}  =\vb{C}+\iu \vb{M}(\omega\mathbf{I}- \vb{\Omega} + \iu \vb{\Gamma} )^{-1}\vb{M}^{\mathsf{T}} .
	\label{eq:SmatCMTmain}
\end{equation}
We use two terms throughout this work. \textit{Bare modes}
are the CMT mode coordinates of \eqref{eq:CMT1main}.
\textit{Dressed modes} are those underpinning the
resonances actually observed in the scattering problem. Each pole of the S-matrix \eqref{eq:SmatCMTmain} is
one dressed mode. One can see that for the case of diagonal $\vb{\Omega} - \iu \vb{\Gamma}$ bare modes and dressed modes coincide with each other. While for a pair of interacting modes, it is possible to link bare and dressed modes, in a general case this connection is rather unknown.
This distinction, though rarely emphasized, is noted in Ref.~\cite{Suh2004}. To the best of our knowledge, a rigorous link between quasi-normal modes and bare CMT modes in the general case remains an open challenge.

Generally, the background non-resonant scattering matrix $\vb{C}$ can have both isotropic and anisotropic contributions. The anisotropic part may produce conversion similar to a birefringent wave-plate. However, such contributions are typically low for thin dielectric membranes, hence we assume the background to be isotropic:
\begin{equation}\label{eq:Cmain}
    \vb{C} \approx
    \begin{pmatrix}
        0&\rho&\tau&0\\
        \rho&0&0&\tau\\
        \tau&0&0&\rho\\
        0&\tau&\rho&0
    \end{pmatrix}.
\end{equation}
In the absence of background dissipation, this matrix is unitary, $\textbf{C}^\dag\textbf{C}=\mathbf{I}$, which determines that $|\rho|^2+|\tau|^2 =1$ and $\arg(\rho) = \arg(\tau)\pm\frac{\pi}{2}$.
It is convenient to parametrize it using two real parameters
\begin{equation}
	\tau=\eu^{\iu\alpha}\cos\beta, \quad
	\rho=\iu \eu^{\iu \alpha}\sin\beta.
    \label{eq:taurho_main}
\end{equation}
where the phase $\alpha\in[0,2\pi]$, and the merit of background transparency $\beta\in[-\pi/2,\pi/2]$.

We now focus on further simplifications of CMT parameters related to the  membrane-metasurface symmetry.
The out-of-plane mirror symmetry of a flat membrane assigns a definite parity $p_n$ to each mode: even ($p_n = +1$) and  odd ($p_n = -1$). This constrains the coupling parameters on opposite sides of the membrane as~\cite{Kumar2025ACSPhot}:
\begin{equation}\label{eq:parity_main}
    m^\prime_{n \mathrm{R}}= p_n\, m_{n \mathrm{L}}, \quad m^\prime_{n \mathrm{L}}= p_n\, m_{n \mathrm{R}}.
\end{equation}
Combined with the time-reversal constraint \eqref{eq:timereversalconstraint} and the isotropic background~\eqref{eq:Cmain}, this yields (Supporting Information, Section~\ref{app:CMTC})
\begin{equation}\label{eq:mRmLequal}
    |m_{n\mathrm{R}}| = |m_{n\mathrm{L}}|,
\end{equation}
for every bare mode irrespective of its parity: each couples equally to LCP and RCP waves, so its far-field is \textit{linearly polarized}. A single resonance therefore cannot generate $\mathrm{CD}_{\mathrm{cross}}$ and at least two bare modes must cooperate.
Note that Eq.~\eqref{eq:mRmLequal} holds true only for bare modes. As we show below, when bare modes overlap in frequency, the off-diagonal entries in $\vb{\Gamma}$ can hybridize them into \textit{dressed eigenmodes} whose far-fields may turn elliptical and provide mode-level chirality. The essential differences between bare and dressed modes are collected in Tab.~\ref{tab:bare_vs_dressed}.

From the energy conservation requirement \eqref{eq:energyconstraint} it follows, regardless of the mode parity, that $\gamma_n = 2 |m_{n \text{R}}|^2 = 2 |m_{n \text{L}}|^2$. Combining this with the parametrization as in Eq.~\eqref{eq:taurho_main} and the time-reversal constraint \eqref{eq:timereversalconstraint}, we can express the coupling parameters in the following form (Supporting Information, Section~\ref{app:CMTC}):
\begin{equation}
\begin{split}
    m_{n \text{R}}     &= \iu^{(1 + p_n)/2} \sqrt{\frac{\gamma_n}{ 2}}  \exp \left(\iu \frac{\alpha + p_n \beta}{2} - \iu \theta_n \right), \\
    m_{n \text{L}}     &= \iu^{(1 + p_n)/2} \sqrt{\frac{\gamma_n}{ 2}}  \exp \left(\iu \frac{\alpha + p_n \beta}{2} + \iu \theta_n \right).
\end{split}
\label{eq:mtheta}
\end{equation}
Here the prefactor $\iu^{(1 + p_n)/2}$ is equal to $\iu$ for $p_n=1$ and to $1$ for $p_n= - 1$. Physically, the angle $\theta_n$ characterizes the direction of the far-field linear polarization of a CMT mode with respect to the $x$-axis.

The mechanisms by which two or more modes produce nonzero $\mathrm{CD}_{\mathrm{cross}}$ depend on whether the interacting modes are \textit{port-orthogonal}. We call two modes $n$ and $\ell$ port-orthogonal when
\begin{equation}
    \vb{m}_{n}^{\dagger}\, \vb{m}_{\ell}  = 0.
    \label{eq:portorthog}
\end{equation}
Non-orthogonal modes coherently drive one another through the shared radiative channels, establishing an indirect coupling mediated by the radiation continuum~\cite{Suh2004}.
By virtue of energy conservation $2\vb{\Gamma} = \vb{M}^{\dagger}\vb{M}$, the port-orthogonality~\eqref{eq:portorthog} is equivalent to $\gamma_{n\ell} = 0$ with $n \neq \ell$: port-orthogonal modes radiate into non-overlapping far-field patterns and evolve independently, each contributing an independent Lorentzian to the S-matrix.

This leads to two distinct mechanisms of cross conversion:
\begin{enumerate}[label=(\Alph*)]
    \item \textit{Port-orthogonal mechanism.} Modes of opposite parity are always port-orthogonal, and the matrix $\mathbf{\Omega} - \iu \mathbf{\Gamma}$ is diagonal. The S-matrix~\eqref{eq:SmatCMTmain} decomposes into a sum of independent Lorentzian contributions. Cross conversion dichroism arises from the interference of these contributions.
    \item \textit{Port-non-orthogonal mechanism.} Same-parity modes with non-orthogonal far-fields hybridize through their shared radiation channels, and the dressed modes can become elliptically polarized. While alone they give $\Delta T_{\text{cross}} = 0$, a nonzero conversion dichroism needs an extra opposite-parity mode or material losses.
\end{enumerate}
We illustrate each mechanism with a minimal model and full-wave numerics. Our findings are related to the theory developed in Ref.~\cite{Sun2026arXiv}.

\subsection{Port-orthogonal mechanism: opposite-parity modes}
\label{sec:example_portorthog}

\begin{figure*}
    \centering
    \includegraphics[width=1\linewidth]{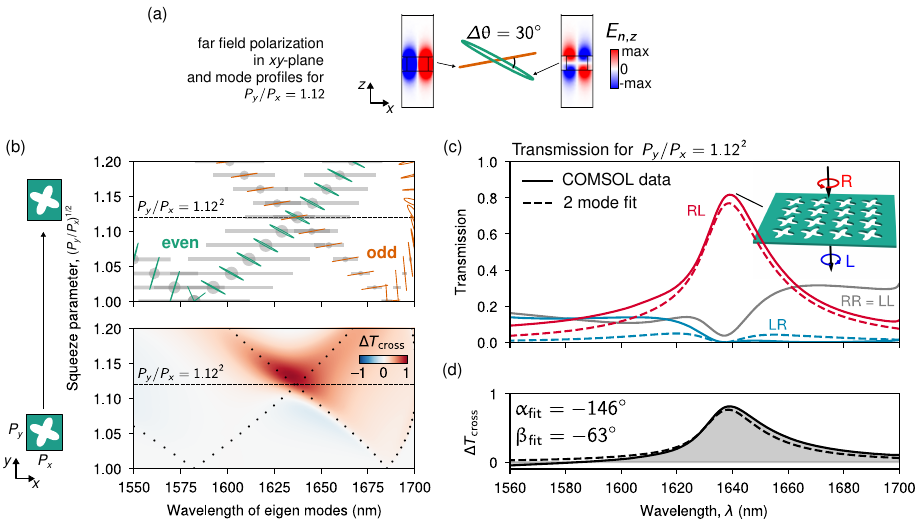}
    \caption{
    \textbf{Port-orthogonal mechanism: opposite-parity modes.}
    (a)~$E_{n,z}$ field profiles of the two modes at the crossing, illustrating their opposite up-down mirror parities.
    (b, top)~Eigenmode wavelengths under area-preserving squeezing of the unit cell ($P_xP_y=\text{const}$), reducing the symmetry from $C_4$ to $C_2$. The $C_4$ double degeneracy splits into two branches; color encodes the parity $p_n=\pm1$~$\left( ^{\text{even}}_{\text{odd}} \right)$, the colored lines show the  far-field polarization, and gray bars mark the linewidth $\Delta\lambda_n=\lambda_n/Q_n$. Opposite-parity modes cross near $\lambda\approx1635\nm$ at $P_y/P_x\approx1.12^2$.
    (b, bottom)~Map of $\Delta T_{\text{cross}}$ in the $(P_y/P_x,\lambda)$ plane from full-wave numerics: the conversion dichroism peaks at the opposite-parity crossing, in agreement with the master equation~\eqref{eq:dT_orthog} which predicts the maximum at $\omega_1=\omega_2$.
    (c,d)~Co- and cross-polarized transmission (c) and cross-conversion dichroism $\Delta T_{\text{cross}}$~\eqref{eq:dTcross} (d) taken at $P_y/P_x=1.12^2$: full-wave numerics versus the two-mode model~\eqref{eq:tLR_orthog}--\eqref{eq:dT_orthog}, with only the background parameters fitted, $\alpha_{\text{fit}}=-146^{\circ}$, $\beta_{\text{fit}}=-63^{\circ}$.
    Simulation parameters: relative permittivity $\varepsilon = 12$, membrane height $H = 400\nm$, fixed $P = \sqrt{P_xP_y} = 1\um$, hole contour from Eq.~\eqref{eq:parametric} with $r_0 = 300\nm$, $r_1 = 123\nm$, and $\theta = \pi/8$.
    }
    \label{fig:squeeze}
\end{figure*}

Opposite-parity modes ($p_1=-p_2$) automatically satisfy the port-orthogonality condition~\eqref{eq:portorthog} via the parity relations~\eqref{eq:parity_main}, regardless of the polarization angles $\theta_n$. The dissipation matrix is therefore diagonal,
\begin{equation}
    \vb{\Omega} - \iu \vb{\Gamma} = \begin{pmatrix}
        \Omega_1 & 0 \\
        0 & \Omega_2
    \end{pmatrix}, \quad \Omega_n = \omega_n - \iu\gamma_n,
    \label{eq:OmegaGammaDiag}
\end{equation}
bare and dressed modes coincide, and the S-matrix~\eqref{eq:SmatCMTmain} reduces to a sum of independent Lorentzians
\begin{equation}
    \vb{S} = \vb{C} - \sum_{n} \frac{\vb{m}_n \vb{m}_n^{\mathsf{T}}}{\iu(\omega - \omega_n) - \gamma_n}.
    \label{eq:Sorthog_main}
\end{equation}
Substituting the parametrization~\eqref{eq:mtheta} gives the cross-polarized amplitudes in closed form:
\begin{equation}
\begin{split}
    &t_{\mathrm{LR}} = \sum_n B_n\,\eu^{-2\iu\theta_n}, \qquad t_{\mathrm{RL}} = \sum_n B_n\,\eu^{+2\iu\theta_n}, \\
    &B_n = \frac{(\gamma_n/2)\,\eu^{\iu(\alpha + p_n\beta)}}{\iu(\omega - \omega_n) - \gamma_n}.
\end{split}
    \label{eq:tLR_orthog}
\end{equation}
For a single resonance $|t_{\mathrm{LR}}^{(n)}| = |t_{\mathrm{RL}}^{(n)}|$, consistent with Eq.~\eqref{eq:mRmLequal}. Cross-conversion dichroism therefore arises only from interference between the two Lorentzians. Denoting the polarization-angle difference $\Delta\theta_{12} = \theta_1-\theta_2$, the chiral signal carries a universal angular factor,
\begin{equation}
    \Delta T_{\text{cross}} = |t_{\mathrm{RL}}|^2 - |t_{\mathrm{LR}}|^2  =  -4\sin(2\Delta\theta_{12})\,\Im\!\left(B_1 B_2^*\right).
    \label{eq:dT_orthog}
\end{equation}
It is possible to show using \eqref{eq:tLR_orthog} that $\left| \Delta T_{\text{cross}}\right| \leq 1$ at degenerate point $\omega_1 = \omega_2$.
Opposite parity puts the two background phases $\eu^{\iu(\alpha+p_n\beta)}$ on opposite branches, so $B_1 B_2^*$ carries a relative factor $\eu^{-2\iu p_1\beta}$, which makes $\Im(B_1 B_2^*)\neq 0$ in general. The cross-conversion dichroism is therefore nonzero whenever $\sin 2\Delta\theta_{12}\neq 0$.%
Chirality is encoded in the relative complex phase between two linearly polarized bare resonances of opposite parity, not at the level of any single eigenmode.

Now we analyze the main result \eqref{eq:dT_orthog} for the two port-orthogonal modes. The maximum values of \eqref{eq:dT_orthog} is achieved for 50\% background transmission ($\beta = \pi/4$), 45 degree angle between far-field linear polarizations, $\Delta\theta_{12} = \pm \pi/4$, and equal real frequencies $\omega_1 = \omega_2$. As soon as decay rates are different $\gamma_1 \neq \gamma_2$, one should observe antisymmetric flip. We illustrate different regimes in Fig.~\ref{fig:AAtheory}.

We now illustrate the theory with a numerical example shown in Fig.~\ref{fig:squeeze}. We start our design with  a freestanding membrane metasurface with the height $H$ and made of lossless dielectric with relative permittivity $\varepsilon$. The membrane has a periodic perforation with the period $P_x$ ($P_y$) along the $x$-axis ($y$-axis) of a four-petal holes defined by a parametric equation
\begin{equation}
    r(\phi) = r_0 + r_1 \cos [4(\phi + \theta)].
    \label{eq:parametric}
\end{equation}
By fixing $P_x = P_y$, we start with a $C_4$ symmetric design which despite being 2D chiral forbids any cross conversion as was derived in Sec.~\ref{sec:selectionrules}. All modes which can radiate into the continuum in this system are double degenerate (this is a consequence of $C_{q}$ symmetry for all systems with $q \geq 3$), see more in Refs.~\cite{Kumar2025ACSPhot,Gorkunov2025AOM,Igoshin2024PRB,CanosValero2024PRRes,Gladyshev2020PRB}. Next, we do the area preserving squeezing of the unit cell by keeping $P_x P_y = \operatorname{const}$ [see Fig.~\ref{fig:squeeze}(b)]. This operation reduces rotational symmetry from four fold to a two fold. Degeneracy of the modes is lifted and we observe a splitting in their positions [shown by color in Fig.~\ref{fig:squeeze}(b)]. The colored lines shows the far-field polarization of the eigenmodes. Since we do not break up-down mirror symmetry, the parity is preserved. Around $1635\nm$ and $P_y / P_x = 1.12^2$ modes of different parities cross. Since these modes are port-orthogonal, and they are spectrally isolated from any other modes, their far polarization is linear and can be described by an angle $\theta_n$ with the $x$-axis. This is in the agreement with Eq.~\eqref{eq:mRmLequal}. The numerical values extracted from the eigen value solver of COMSOL are $\Omega_1 \approx  (1150-\iu20)\times10^{12}$\,rad/s, $\Omega_2 \approx  (1152-\iu8)\times10^{12}$\,rad/s, $\theta_1 \approx -0.39$, $\theta_2 \approx 0.14$, $p_1 = +1$, and $p_2 = -1$ [field profiles in Fig.~\ref{fig:squeeze}(a)]. Since bare modes for port-orthogonal crossing coincide with the true dressed modes, we are left with only two free parameters in Eqs.~\eqref{eq:tLR_orthog}--\eqref{eq:dT_orthog}: the background parameters $\alpha$ and $\beta$ introduced in Eq.~\eqref{eq:taurho_main}. By varying only these two parameters we are able to achieve good agreement between full numerical transmission calculations and the two mode model at $\alpha_{\text{fit}} = - 146^{\circ}$ and $\beta_{\text{fit}} = - 63^{\circ}$ [see Fig.~\ref{fig:squeeze}(c,d)]. Scanning $\Delta T_{\text{cross}}$ over the full $(P_y/P_x,\lambda)$ plane [Fig.~\ref{fig:squeeze}(b)] confirms that the conversion dichroism \eqref{eq:dTcross} peaks precisely at the opposite-parity crossing, as predicted by the master equation~\eqref{eq:dT_orthog} which reaches its maximum at  $\omega_1=\omega_2$.

\subsection{Port-non-orthogonal mechanism: same-parity modes}
\label{sec:example_portnonorthog}

\begin{figure*}
    \centering
    \includegraphics[width=\linewidth]{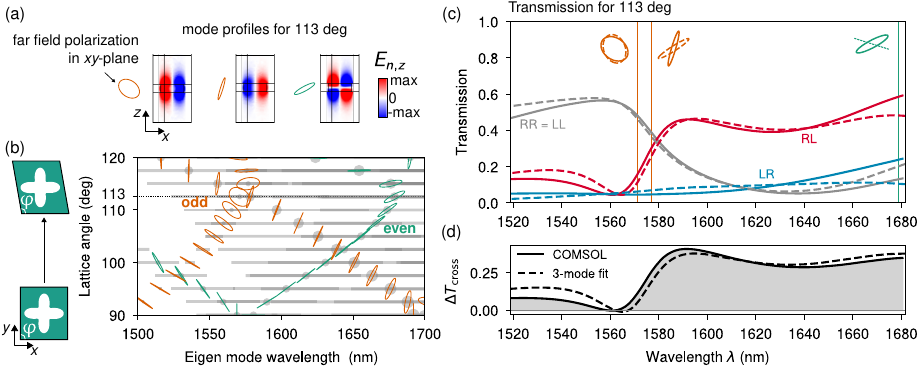}
    \caption{
    \textbf{Port-non-orthogonal mechanism: same-parity modes.}
    (a)~$E_{n,z}$ field profiles of the three involved modes at the crossing, illustrating their up-down mirror parities.
    (b)~Eigenmode wavelengths versus the lattice angle $\varphi$, deforming a rectangular lattice ($\varphi=90^\circ$) into a 2D-chiral monoclinic one. Color encodes the parity $p_n=\pm1$ $\left( ^{\text{even}}_{\text{odd}} \right)$, ellipses show the far-field polarization, and gray bars mark the linewidth $\Delta\lambda_n=\lambda_n/Q_n$. Near $\varphi\approx113^\circ$ two same-parity modes cross; port-non-orthogonal mixing turns one dressed mode circularly polarized, violating the bare-mode result~\eqref{eq:mRmLequal}.
    (c,d)~Co- and cross-polarized transmission (c) and cross-conversion dichroism $\Delta T_{\text{cross}}$~\eqref{eq:dTcross} (d): full-wave numerics versus the three-mode model~\eqref{eq:SmatCMTmain}, fitted with the parameters listed in Tab.~\ref{tab:mode_comparison}.
    Simulation parameters: $\varepsilon=12$, height $H=300\nm$, hole contour from Eq.~\eqref{eq:parametric} with $r_0=300\nm$, $r_1=123\nm$, $\theta=0\,\text{deg}$; periods $P_x=1\um$, $P_y=1.05\um$.
    }
    \label{fig:port-non-orthogonal}
\end{figure*}

Now we turn to the analysis of the port-non-orthogonal mechanism of the modes with close frequencies. In the current CMT model, we always have $\Delta T_{\text{cross}} = 0$ for 2 and more  modes of the same parity spectrally isolated from modes of an opposite parity. Either a mode of opposite parity or the presence of material losses is always required (Supporting Information, Section~\ref{app:SmatPortNonOrthog}).

For that purpose we consider a three mode model with a non-diagonal elements in the $\vb{\Gamma}$ matrix. In that case bare CMT modes do not coincide with the true eigenmodes of the system and diagonalization procedure must be done. In general, analytics becomes cumbersome and heavy to analyse, yet we still present a convenient parametrization approach.
Substituting the parametrization~\eqref{eq:mtheta} together with the parity relations~\eqref{eq:parity_main} into the port-orthogonality condition~\eqref{eq:portorthog} gives, for two same-parity modes ($p_1=p_2$),
\begin{equation}
    \vb{m}_1^{\dagger}\vb{m}_2  =  2\sqrt{\gamma_1\gamma_2} \, \cos(\Delta\theta_{12}),\qquad \Delta\theta_{12}=\theta_1-\theta_2,
    \label{eq:M1M2_sameparity}
\end{equation}
and $\vb{m}_1^{\dagger}\vb{m}_2 = 0$ for opposite parity ($p_1= - p_2$).
Same-parity modes are therefore port-orthogonal if and only if  their far-field polarizations are perpendicular [$\cos(\Delta\theta_{12})=0$]. However,  generally $\cos\Delta\theta_{12}\neq 0$ and the modes are port-non-orthogonal. Energy conservation $2\vb{\Gamma}=\vb{M}^{\dagger}\vb{M}$ then identifies the off-diagonal decay rate as $\gamma_{12} = \sqrt{\gamma_1\gamma_2}\,\cos (\Delta\theta_{12})$. We can generalize it for 3 mode case as
\begin{equation}
    \gamma_{ij} = \frac{1 + p_i p_j}{2}\sqrt{\gamma_i\gamma_j}\,\cos(\Delta\theta_{ij}),
    \label{eq:gammaijmain}
\end{equation}
and the CMT matrix $\vb{\Omega} - \iu \vb{\Gamma}$ \eqref{eq:OmegaGamma} acquires off-diagonal elements.

Using general equation for the scattering matrix \eqref{eq:SmatCMTmain} and parametrization \eqref{eq:mtheta} we construct a fitting model based on several meaningful parameters: two parameters to define amplitude and phase of the transmission, e.g. angles $\alpha \in [0, 2\pi]$ and $\beta \in [-\pi/2, \pi/2]$; and three parameters for each bare mode, namely its resonant position $\omega_n$, spectral linewidth $\gamma_n$, and its far-field polarization angle $\theta_n$.
Once it is done, the complex orthogonal diagonalization matrix $\vb{T}$ ($\vb{T}^{\mathsf{T}} \vb{T} = \mathbf{I}$) can be found numerically which gives the frequency of the dressed modes $\tilde{\Omega}_n$:
\begin{equation}
    \vb{T}^{\mathsf{T}} (\vb{\Omega} - \iu \vb{\Gamma}) \vb{T} = \begin{pmatrix}
        \tilde{\Omega}_1 & 0 & 0 \\
        0 & \tilde{\Omega}_2 & 0 \\
        0 & 0 & \tilde{\Omega}_3
    \end{pmatrix},
    \label{eq:diagonal3mode}
\end{equation}
and the corresponding dressed coupling parameters
\begin{equation}
    \tilde{\vb{M}} = \vb{M} \vb{T}.
    \label{eq:Mtilde}
\end{equation}
Time reversal constraint \eqref{eq:timereversalconstraint} and energy conservation identity \eqref{eq:energyconstraint} are no longer satisfied for the dressed parameters $\tilde{\vb{M}}$ defined by the transformation $\vb{T}$ in Eq.~\eqref{eq:Mtilde}. Hence, conclusion about linear polarization \eqref{eq:mRmLequal} is no longer applicable to the dressed modes, and \textit{elliptically polarized modes are allowed}. Finally, once the $S$-matrix  is calculated numerically via Eq.~\eqref{eq:SmatCMTmain}, the cross conversion dichroism can be calculated directly using its connection with the transmission coefficients, Eq.~\eqref{eq:Smatrix}.

Let us demonstrate a vivid example of such 3 mode mixing, which produce circularly polarized in the far-field modes.
We start with a free standing membrane made of dielectric with $\varepsilon= 12$ and height $H = 300\nm$. It has a perforation of four-fold holes defined by a parametric equation~\eqref{eq:parametric} with  $r_0 = 300\nm$, $r_1 = 123\nm$, and $\theta = 0$. The holes form a rectangular arrangement with the periods $P_x = 1\um$ and $P_y = 1.05\um$. Since the four-fold cut is aligned with the periodic arrangement, there is no 2D chirality yet~\cite{Sinev2025NC}. Next, we deform the lattice angle $\varphi$ from $90^\circ$ to larger values, while keeping other parameters constant [Fig.~\ref{fig:port-non-orthogonal}(b)]. In doing so, we transition from rectangular lattice to a 2D chiral monoclinic arrangement, which lacks all possible in-plane mirror symmetries~\cite{Toftul2024PRL,Sinev2025NC,Toftul2025Nanophotonics,Sun2025SA}.

\begin{table}
  \centering
  \caption{\textbf{Three mode model fitting parameters.} All complex angular frequencies are given in units of $10^{12}$\,rad/s. Columns from left to right: the mode parity $p_n$; the eigenfrequencies $\Omega_n^{\text{COMSOL}} = 2\pi f_n$ of the eigenmode solver, where $f_n$ is the COMSOL eigenfrequency conjugated to the $\eu^{-\iu\omega t}$ convention; the dressed-mode frequencies $\tilde{\Omega}_n$ obtained by the diagonalization~\eqref{eq:diagonal3mode} of the fitted model; the fitted bare CMT mode frequencies $\Omega_n$ and polarization angles $\theta_n$. The fitted background angles are $\alpha = 13.8^\circ$ and $\beta = -55.3^\circ$.}
  \label{tab:mode_comparison}
  \begin{tabular*}{\linewidth}{@{\extracolsep{\fill}}ccccc}
    \toprule
    $p_n$ & $\Omega_n^{\text{COMSOL}}$ & $\tilde{\Omega}_n$ (dressed) & $\Omega_n$ (fit) & $\theta_n$ (fit) \\
    \midrule
    $+1$ & $1121.9 - \iu 36.2$ & $1121.9 - \iu 36.1$ & $1121.9 - \iu 36.1$ & $16.9^\circ$ \\
    $-1$ & $1194.4 - \iu 41.5$ & $1194.4 - \iu 41.5$ & $1185.6 - \iu 29.8$ & $166.8^\circ$ \\
    $-1$ & $1198.8 - \iu 12.1$ & $1198.9 - \iu 12.1$ & $1207.7 - \iu 23.9$ & $119.3^\circ$ \\
    \bottomrule
  \end{tabular*}
\end{table}

In Fig.~\ref{fig:port-non-orthogonal}(b) we show the eigenmode evolution as a function of the lattice angle, where color shows the up-down mirror parity (illustrated by the field profiles in panel (a)) and small ellipses show the far-field polarization. Gray horizontal bars show spectral linewidth defined as $\Delta \lambda_n = \lambda_n / Q_n$. Around $\varphi = 113^{\circ}$ we observe a crossing of the same parity modes. One of the modes is circularly polarized. This is the consequence of the port-non-orthogonal mixing, which violates the result \eqref{eq:mRmLequal} obtained for the bare modes, as numerical eigenvalue solver provides rather dressed modes.
This crossing is not spectrally isolated and in its vicinity there a third mode of opposite parity. Hence, to describe the cross conversion dichroism in this case we have to employ the three mode model developed in this section.

We calculate the transmission and conversion dichroism using full numerical model [Fig.~\ref{fig:port-non-orthogonal}(c,d)]. Next, we use a numerical fit of the three mode model. The fitting parameters are modes bare frequencies $\Omega_n$ and polarization angles $\theta_n$, as well as the background angles $\alpha$ and $\beta$ from \eqref{eq:taurho_main}. We set the target of the minimization procedure to be the minimal difference in the co- and cross-polarized transmissions, as well as the closeness of the dressed frequencies to the ones obtained by a numerical solver. We show the final values in the Tab.~\ref{tab:mode_comparison}, and observe a good agreement between the full numerical model and a three mode CMT model.

\subsection{Role of losses}
\label{sec:losses}

\begin{figure}
    \centering
    \includegraphics[width=\linewidth]{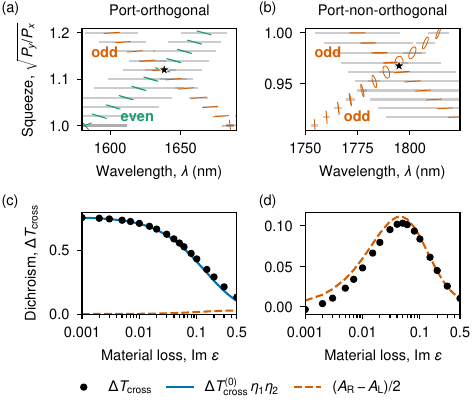}
    \caption{
    \textbf{Material losses act in opposite ways on the two mechanisms.}
    (a,b)~Eigenmode wavelengths versus the squeeze parameter $\sqrt{P_y/P_x}$ at $P_xP_y = P^2$. Color encodes the parity, ellipses show the far-field polarization, and gray bars mark the linewidth $\lambda_n/Q_n$. The star marks the point of the transmission calculation.
    (a)~Port-orthogonal crossing of Fig.~\ref{fig:squeeze}. Star: $\sqrt{P_y/P_x} = 1.12$, $\lambda \approx 1635\nm$.
    (b)~Port-non-orthogonal crossing of two odd modes. Free-standing membrane with $\varepsilon = 12 + \iu \Im \varepsilon$, $H = 289\nm$, $P = 1\um$, and one rectangular hole of $222 \times 556\nm$ per cell, rotated by $\pi/8$ from the $y$-axis. Star: $\sqrt{P_y/P_x} = 0.968$, $\lambda = 1794.9\nm$.
    (c,d)~Conversion dichroism $\Delta T_{\text{cross}}$~\eqref{eq:dTcross} at the star wavelength versus the material loss $\Im\varepsilon$ (points), and the circular dichroism in absorption $(A_{\mathrm{R}} - A_{\mathrm{L}})/2$ (dashed), which equals $\Delta T_{\text{cross}}$ for same-parity modes [Eq.~\eqref{eq:dT_abs_main}].
    The solid line in (c) is Eq.~\eqref{eq:loss_orthog_main} with $\eta_n = \gamma_n/(\gamma_n + \gamma_n^{\mathrm{abs}})$ and no free parameter. The radiative rates $\gamma_1 = 19.8\times10^{12}$\,rad/s and $\gamma_2 = 8.0\times10^{12}$\,rad/s and the absorption rates $\gamma_n^{\mathrm{abs}} \approx \kappa_n\Im\varepsilon$, with $\kappa_1 = 37.4\times10^{12}$\,rad/s and $\kappa_2 = 44.4\times10^{12}$\,rad/s, come from the eigenmode calculation.
    In (d), the small difference between the points and the dashed line comes from the nearest even mode, outside the window of (b).
    }
    \label{fig:losses}
\end{figure}

Absorption removes energy from a mode but does not feed an outgoing port.
We add it through the substitution $\vb{\Omega}-\iu\vb{\Gamma}\to\vb{\Omega}-\iu \left( \vb{\Gamma} + \vb{\Gamma}_{\mathrm{abs}}  \right)$, where $\vb{\Gamma}_{\mathrm{abs}}=\operatorname{diag}(\gamma_1^{\mathrm{abs}},\gamma_2^{\mathrm{abs}},\dots)$.
The matrix $\vb{M}$ of coupling to the radiation continuum does not change in the first approximation.
Thus $\gamma_n$ remains the radiative decay rate of mode $n$, and its total decay rate is $\gamma_n + \gamma_n^{\mathrm{abs}}$.
Absorption has opposite effects on the two mechanisms of conversion dichroism:
\begin{enumerate}[label=\textit{(\arabic*)}]
	\item Absorption always weakens $\Delta T_{\text{cross}}$ in the port-orthogonal mechanism.
	\item Absorption is necessary for $\Delta T_{\text{cross}}$ in the port-non-orthogonal mechanism, with a maximum at critical coupling.
\end{enumerate}

\subsubsection{Losses in the port-orthogonal mechanism}

In the port-orthogonal mechanism, absorption reduces the amplitude of each independent Lorentzian term of the S-matrix~\eqref{eq:Sorthog_main}.
At the crossing point and on resonance, $\omega = \omega_1 = \omega_2$, we find (Supporting Information, Section~\ref{app:losses})
\begin{equation}
    \Delta T_{\text{cross}} = \eta_1\eta_2\,\Delta T_{\text{cross}}^{(0)},
    \label{eq:loss_orthog_main}
\end{equation}
where $\Delta T_{\text{cross}}^{(0)}$ is the lossless value, and $\eta_n = \gamma_n/(\gamma_n + \gamma_n^{\mathrm{abs}})$ is the \textit{radiative efficiency}. Losses therefore only weaken this mechanism, and a lossless membrane is optimal. The full-wave results for the crossing of Fig.~\ref{fig:squeeze} follow Eq.~\eqref{eq:loss_orthog_main} [Fig.~\ref{fig:losses}(c)], and the circular dichroism in absorption stays negligible.

\subsubsection{Losses in the port-non-orthogonal mechanism}

Without losses, two spectrally isolated same-parity modes give $\Delta T_{\text{cross}} = 0$, even when a dressed mode is circularly polarized (Supporting Information, Section~\ref{app:SmatPortNonOrthog}). With losses, energy conservation gives (Supporting Information, Section~\ref{app:losses})
\begin{equation}
	\Delta T_{\text{cross}} = (A_{\mathrm{R}} - A_{\mathrm{L}})/2,
	\label{eq:dT_abs_main}
\end{equation}
where $A_{\mathrm{R}}$ ($A_{\mathrm{L}}$) is the absorbed fraction of the incident RCP (LCP) power. The up-down mirror symmetry limits this absorption. The incident wave is the sum of an even and an odd part under $z \to -z$, and each part carries one half of the incident power. Same-parity modes couple to only one part, so $0 \leq A_{\mathrm{R,L}} \leq 1/2$ and
\begin{equation}
    |\Delta T_{\text{cross}}| \leq \frac{1}{4}.
    \label{eq:dT_abs_bound_short}
\end{equation}
To reach the bound, the structure needs $A_{\mathrm{L}} = 0$, that is, a circularly polarized dressed mode, which requires at least two interacting same-parity modes. It also needs $A_{\mathrm{R}} = 1/2$, that is, critical coupling, where the absorption rate equals the radiative rate. Thus this mechanism has an optimum at a finite loss [Fig.~\ref{fig:losses}(d)]. There, $(A_{\mathrm{R}} - A_{\mathrm{L}})/2$ follows $\Delta T_{\text{cross}}$, and the small difference comes from a nearby mode of opposite parity.
Thus the bound for the port-non-orthogonal mechanism is four times lower than for the port-orthogonal one, which allows $|\Delta T_{\text{cross}}| = 1$ [Eq.~\eqref{eq:dT_orthog}].

\section{Conclusion}

We have analyzed, both numerically and theoretically, the manifestation of chiral properties of single-layer dielectric structured membrane metasurfaces. We have assumed that the metasurface is vertically symmetric, but it has a broken in-plane symmetry introduced by deformations of the unit cell.
A substrate or non-ideal fabrication breaks the vertical mirror symmetry~\cite{Gorkunov2025AOM,Kumar2025ACSPhot,Glebov2026arXiv}, and then $\mathrm{CD}_{\mathrm{co}}$ generally coexists with $\Delta T_{\mathrm{cross}}$. Our parity-based results, including the bound $|\Delta T_{\mathrm{cross}}|\leq 1/4$ for port-non-orthogonal crossing, hold only for exact vertical symmetry.
Although for a long time chiral optics was associated with objects fully lacking mirror symmetry planes, nowadays, those retaining a horizontal out-of-plane symmetry (so-called 2D chiral) attract increasing attention and demonstrate remarkably strong chiral optical effects.
We have explicitly demonstrated the main principles which reveal how single-layer metasurfaces can demonstrate strong resonant chiral effects.
As a matter of fact, strong circular dichroism appears as a cross-polarization conversion due to resonances of photonic eigenmodes in a finite-width metasurface. We have explored this effect for a specific example of a $C_4$ rotation-symmetric metasurface with in-plane perturbations. Breaking the in-plane mirror symmetry and lifting the rotation symmetry unlocks circular conversion dichroism near resonances.  We have established general selection rules for this effect and have traced its connection with the eigenmode interference and inter-coupling by employing the chiral coupled-mode theory.
Finally, we presented a comprehensive analysis of the role of losses in conversion dichroism in similar systems.

\section*{Acknowledgments}
The authors gratefully acknowledge M. Gorkunov, whose contributions were central to the theory development of this work.
B.K. acknowledges a support of Prime Minister Research Fellowship from Ministry of Science and Technology, India.  Y.K. thanks Dr. Peng Xie for his initial participation in a related project.
The collaborative work on this project began in 2025.

\bibliographystyle{unsrt}
\bibliography{ref}

\newpage  %

\widetext

\vspace{14pt}
\begin{center}
    {\Large\bfseries Supporting Information}
\end{center}
\vspace{8pt}

\appendix
\setcounter{section}{0}
\renewcommand{\thesection}{S\arabic{section}}
\makeatletter
\def\@sectioncntformat#1{\csname the#1\endcsname.}
\def\@hangfrom@section#1#2#3{#1\@if@empty{#2}{#3}{#2\@if@empty{#3}{}{\ #3}}}
\makeatother
\counterwithout{equation}{section}
\setcounter{equation}{0}
\setcounter{figure}{0}
\setcounter{table}{0}
\renewcommand{\theequation}{S\arabic{equation}}
\renewcommand{\thefigure}{S\arabic{figure}}
\renewcommand{\thetable}{S\arabic{table}}
\renewcommand{\theHequation}{S\arabic{equation}}
\renewcommand{\theHfigure}{S\arabic{figure}}
\renewcommand{\theHtable}{S\arabic{table}}

\section{Coupled-mode theory of flat membranes}

\label{app:CMTC}

Setting $\dd\boldsymbol{\psi}/\dd t = 0$ in Eq.~\eqref{eq:CMT1main} gives the steady-state amplitudes $\boldsymbol{\psi} = \iu(\omega\mathbf{I} - \vb{\Omega} + \iu\vb{\Gamma})^{-1}\vb{M}^{\mathsf{T}}\vb{a}$. Substituting them into Eq.~\eqref{eq:CMT2main} yields the scattering matrix~\eqref{eq:SmatCMTmain}. Throughout this section we abbreviate the effective non-Hermitian Hamiltonian as $\vb{H} \equiv \vb{\Omega} - \iu\vb{\Gamma}$, so that $\vb{S} = \vb{C} + \iu\vb{M}(\omega\mathbf{I} - \vb{H})^{-1}\vb{M}^{\mathsf{T}}$.

\subsection{Coupling parameters of bare modes}

A flat membrane is invariant with respect to the $z\leftrightarrow -z$ interchange and its eigenstates can be categorized by their parity with respect to the $z$-axis. We denote even modes having even transversal and odd normal electric field components (see the derivation using group theory in the SM of Ref.~\cite{Kumar2025ACSPhot}):
\begin{align}\label{even}
\begin{split}
	E_{\mathrm{even}\ x,y}(x,y,z)&=E_{\mathrm{even}\ x,y}(x,y,-z),\\
    E_{\mathrm{even}\ z}(x,y,z)&=-E_{\mathrm{even}\ z}(x,y,-z),
\end{split}
\end{align}
while the opposite is true for the odd modes:
\begin{align}\label{odd}
\begin{split}
	E_{\mathrm{odd}\ x,y}(x,y,z)&=-E_{\mathrm{odd}\ x,y}(x,y,-z)\\%E_{{\rm e}\ x,y}(x,y,-z),\\
    E_{\mathrm{odd}\ z}(x,y,z)&=E_{\mathrm{odd}\ z}(x,y,-z).
\end{split}
\end{align}
For the coupling parameters, this allows relating those on the opposite metasurface sides as
\begin{equation}\label{eq:meven}
	m'_{\mathrm{even}\,\mathrm{R}}=m_{\mathrm{even}\,\mathrm{L}},\ m'_{\mathrm{even}\,\mathrm{L}}=m_{\mathrm{even}\,\mathrm{R}},
\end{equation}
for even modes, and
\begin{equation}\label{eq:modd}
	m'_{\mathrm{odd}\,\mathrm{R}}=-m_{\mathrm{odd}\,\mathrm{L}},\ m'_{\mathrm{odd}\,\mathrm{L}}=-m_{\mathrm{odd}\,\mathrm{R}},
\end{equation}
for the odd ones. In the parity notation of the main text these are $p_n=+1$ (even) and $p_n=-1$ (odd), so Eqs.~\eqref{eq:meven} and~\eqref{eq:modd} are the two cases of Eq.~\eqref{eq:parity_main}.

For adequate description of chiral effects, we have to express all polarization conversion effects by the participating modes. The background channels are then polarization independent and amount to an isotropic background S-matrix:
\begin{equation}\label{eq:Cmatr}
	\vb{C}=
	\begin{pmatrix}
		0 & \rho & \tau & 0\\
		\rho & 0 & 0 & \tau\\
		\tau & 0 & 0 & \rho\\
		0 & \tau & \rho & 0
	\end{pmatrix}.
\end{equation}
It can be set unitary by parameterizing
\begin{equation}\label{eq:taurho}
	\tau=\eu^{\iu\alpha}\cos\beta,\qquad
    \rho=\iu\eu^{\iu\alpha}\sin\beta.
\end{equation}
with a pair of real parameters: the phase $\alpha\in[0,2\pi]$, and the merit of background transparency $\beta\in[-\pi/2,\pi/2]$.

Then, by substituting Eqs.~\eqref{eq:meven} and \eqref{eq:Cmatr} into time reversal constraint $\vb{C} \vb{M}^{*} = - \vb{M}$ [Eq.~\eqref{eq:timereversalconstraint} in the main text] one obtains for an even eigenstate:
\begin{equation}
	m_{\mathrm{even}\,\mathrm{R}}=-m_{\mathrm{even}\,\mathrm{L}}^*(\tau+\rho),\
	m_{\mathrm{even}\,\mathrm{L}}=-m_{\mathrm{even}\,\mathrm{R}}^*(\tau+\rho),
\end{equation}
while for an odd one, similarly using Eq.~\eqref{eq:modd} yields:
\begin{equation}
	m_{\mathrm{odd}\,\mathrm{R}}=m_{\mathrm{odd}\,\mathrm{L}}^*(\tau-\rho),\
	m_{\mathrm{odd}\,\mathrm{L}}=m_{\mathrm{odd}\,\mathrm{R}}^*(\tau-\rho).
\end{equation}
Substituting here the parameterization \eqref{eq:taurho} yields:
\begin{equation}\label{eq:m1Rm1Leq}
	m_{\mathrm{even}\,\mathrm{L}}=-m_{\mathrm{even}\,\mathrm{R}}^*\eu^{\iu(\alpha+\beta)},
\end{equation}
\begin{equation}\label{eq:m2Rm2Leq}
	m_{\mathrm{odd}\,\mathrm{L}}=m_{\mathrm{odd}\,\mathrm{R}}^*\eu^{\iu(\alpha-\beta)}.
\end{equation}
Hereby we obtain an important very general restriction of the coupling parameters of modes of a planar membrane:
\begin{equation}\label{eq:mLmR}
    |m_{\mathrm{even,odd}\,\mathrm{R}}|=|m_{\mathrm{even,odd}\,\mathrm{L}}|,
\end{equation}
i.e., regardless of their parity, they are coupled equally strongly to LCP and RCP waves.

Next, taking the diagonal elements of the matrix equation \eqref{eq:energyconstraint}, allows relating the absolute values of the coupling parameters with the  diagonal elements of the matrix $\vb{\Gamma}$ regardless of the mode parity:
\begin{equation}\label{eq:gamman}
	\gamma_{n}=2|m_{n\mathrm{R}}|^2=2|m_{n\mathrm{L}}|^2.
\end{equation}

Finally, to express the complex phases, one can introduce the angles  $\theta_{\mathrm{even,odd}}$ and satisfy Eqs.~\eqref{eq:m1Rm1Leq}--\eqref{eq:gamman} by setting:
\begin{equation}\label{eq:mRLeven}
	m_{\mathrm{even}\,\mathrm{R}}=\iu\sqrt{\frac{\gamma_{\mathrm{even}}}{2}}\eu^{\iu\frac{\alpha+\beta}{2}- \iu\theta_{\mathrm{even}}},\
	m_{\mathrm{even}\,\mathrm{L}}=\iu\sqrt{\frac{\gamma_{\mathrm{even}}}{2}}\eu^{\iu\frac{\alpha+\beta}{2}+ \iu\theta_{\mathrm{even}}},
\end{equation}
\begin{equation}\label{eq:mRLodd}
	m_{\mathrm{odd}\,\mathrm{R}}=\sqrt{\frac{\gamma_{\mathrm{odd}}}{2}}\eu^{\iu\frac{\alpha-\beta}{2}- \iu\theta_{\mathrm{odd}}},\
	m_{\mathrm{odd}\,\mathrm{L}}=\sqrt{\frac{\gamma_{\mathrm{odd}}}{2}}\eu^{\iu\frac{\alpha-\beta}{2}+ \iu\theta_{\mathrm{odd}}}.
\end{equation}
Physically, the angles $\theta_{\mathrm{even,odd}}$ characterize the direction of the far-field polarization of the modes, as they actually are the angles between the linearly polarized electric field  and the $x$-axis.

\section{Port-orthogonal modes (opposite parity)}
\label{app:PortOrthog}

In some particular situations, the corresponding Hamiltonian $\vb{H}$ can be chosen diagonal.
The general precondition for this is the diagonal $\vb{\Gamma}$ in Eq.~\eqref{eq:energyconstraint}, which requires that any pair of modes $n_1$ and $n_2$ has to obey:
\begin{equation}\label{eq:orthog}
m_{n_1\mathrm{R}}m^*_{n_2\mathrm{R}}+m_{n_1\mathrm{L}}m^*_{n_2\mathrm{L}}+m'_{n_1\mathrm{R}}m'^*_{n_2\mathrm{R}}+m'_{n_1\mathrm{L}}m'^*_{n_2\mathrm{L}}=0.
\end{equation}
We will refer to this as a port-orthogonality condition.

It is easy to notice that by virtue of Eqs.~\eqref{eq:meven} and \eqref{eq:modd}, it is always fulfilled for the modes of different parity.
For modes of the same parity, substituting Eqs.~\eqref{eq:mRLeven} or \eqref{eq:mRLodd} reduces Eq.~\eqref{eq:orthog} to
\begin{equation}
    \cos(\theta_{n_1}-\theta_{n_2})=0
\end{equation}
i.e., to the orthogonality of their linear polarizations in the far-field.

Apparently, the total number of port-orthogonal modes is restricted by the number of S-matrix channels, i.e., there can be maximum 2 odd and 2 even modes mutually orthogonal in the far-field. As the corresponding Hamiltonian is diagonal, the matrix inversion in Eq.~\eqref{eq:SmatCMTmain} is trivial and one obtains:
\begin{equation}
     \vb S =\vb C -\sum_{n=1}^{N}\frac{1}{\iu(\omega-\omega_n)-\gamma_n}
    \begin{pmatrix}
        m_{n \mathrm{R}}^2 & m_{n \mathrm{R}}m_{n \mathrm{L}} & m_{n \mathrm{R}}m^\prime_{n \mathrm{R}} & m_{n \mathrm{R}}m^\prime_{n \mathrm{L}}\\
        m_{n \mathrm{L}}m_{n \mathrm{R}} & m_{n \mathrm{L}}^2 & m_{n \mathrm{L}}m^\prime_{n \mathrm{R}} & m_{n \mathrm{L}}m^\prime_{n \mathrm{L}}\\
        m^\prime_{n \mathrm{R}}m_{n \mathrm{R}}&m^\prime_{n \mathrm{R}}m_{n \mathrm{L}} & {m^\prime_{n \mathrm{R}}}^2 & m^\prime_{n \mathrm{R}}m^\prime_{n \mathrm{L}}\\
        m^\prime_{n \mathrm{L}}m_{n \mathrm{R}}&m^\prime_{n \mathrm{L}}m_{n \mathrm{L}}  & m^\prime_{n \mathrm{R}}m^\prime_{n \mathrm{L}}& {m^\prime_{n \mathrm{L}}}^2.
    \end{pmatrix} \label{eq:Sorthog}
\end{equation}

\subsection{Two-mode regime map}
\label{sec:appendix_twomode}

\begin{figure}[h]
    \centering
    \includegraphics[width=0.5\linewidth]{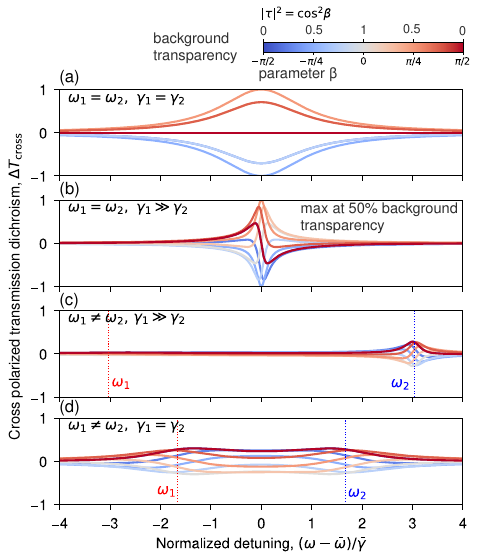}
    \caption{
    \textbf{Cross-conversion dichroism of two opposite-parity resonances.}
    Two-mode model computed from Eq.~\eqref{eq:dT_orthog}. Dimensionless parameters are detuning $(\omega-\bar{\omega})/\bar{\gamma}$ with $\bar{\omega}=(\omega_1+\omega_2)/2$ and $\bar{\gamma}=(\gamma_1+\gamma_2)/2$. Each curve corresponds to a different background transmission $|\tau|^2=\cos^2\!\beta$ (colorbar). Four representative parameter regimes are shown:
    (a)~degenerate frequencies and equal linewidths ($\omega_1=\omega_2$, $\gamma_1=\gamma_2$);
    (b)~degenerate frequencies with mismatched linewidths ($\gamma_1\gg\gamma_2$);
    (c)~non-degenerate frequencies with mismatched linewidths; and
    (d)~non-degenerate frequencies with equal linewidths. Vertical dotted lines in (c) and (d) mark the individual resonance positions $\omega_1$ (red) and $\omega_2$ (blue).}
    \label{fig:AAtheory}
\end{figure}

Figure~\ref{fig:AAtheory} maps the cross-conversion dichroism~\eqref{eq:dT_orthog} of two opposite-parity ($p_2 = -p_1$) resonances across the representative regimes of frequency detuning and linewidth mismatch discussed in Sec.~\ref{sec:example_portorthog}.

Three limits organize the panels of Fig.~\ref{fig:AAtheory}. For degenerate modes of equal linewidth [$\omega_1=\omega_2 = \omega_0$, $\gamma_1=\gamma_2=\gamma$, panel (a)] the dispersive term cancels and Eq.~\eqref{eq:dT_orthog} collapses to a single Lorentzian,
\begin{equation}
    \Delta T_{\mathrm{cross}} = -\operatorname{sign}(p_1)\frac{\gamma^2}{\gamma^2+(\omega - \omega_0)^2}\,\sin(2\beta) \sin(2\Delta\theta_{12}).
    \label{eq:dT_degenerate}
\end{equation}
It is symmetric about resonance and bounded by $|\Delta T_{\mathrm{cross}}|\leq 1$. The bound is reached at $\Delta\theta_{12}=\pm\pi/4$ and $\beta=\pm\pi/4$, a half-transparent background, $|\tau|^2=1/2$.

Keeping the modes degenerate but mismatching their linewidths [$\omega_1=\omega_2=\omega_0$, $\gamma_1\gg\gamma_2$, panel (b)] turns the response into a dispersive \emph{Fano} lineshape, reaching $|\Delta T_{\mathrm{cross}}|=1$ at the half-transparent background $\beta=\pm\pi/4$.

Lifting the frequency degeneracy [panels (c) and (d)] resolves the two resonances at $\omega_1$ and $\omega_2$ (vertical dotted lines): the response splits into two features centered at $\omega_1$ and $\omega_2$, each carrying the dispersive weight $\Im(B_1 B_2^{*})$ of Eq.~\eqref{eq:dT_orthog} with overall sign $-\operatorname{sign}(p_1)\sin(2\Delta\theta_{12})$. They merge into the single Lorentzian of panel~(a) once the splitting drops below $\bar{\gamma}$.

\section{Port-non-orthogonal modes (same parity)}
\label{app:SmatPortNonOrthog}

Consider a pair of modes of the same parity and non-orthogonal far-fields. For definiteness, we choose them to be even and denote them as $1$ and $2$, writing their eigenfrequencies as $\Omega_{1,2}=\omega_{1,2}-\iu\gamma_{1,2}$ and the coupling parameters as
\begin{equation}\label{eq:mRLeven12}
	m_{1,2\,\mathrm{R}}=\iu\sqrt{\frac{\gamma_{1,2}}{2}}\eu^{\iu\frac{\alpha+\beta}{2}- \iu\theta_{1,2}},\
	m_{1,2\,\mathrm{L}}=\iu\sqrt{\frac{\gamma_{1,2}}{2}}\eu^{\iu\frac{\alpha+\beta}{2}+ \iu\theta_{1,2}},
\end{equation}
where the angles $\theta_{1,2}$ characterize the far-field directions.
Substituting these coupling parameters into the off-diagonal element of the condition \eqref{eq:energyconstraint} and using the parity \eqref{eq:meven} we obtain:
\begin{equation}
\label{eq:gamma12}
\gamma_{12}=\gamma_{21}=\sqrt{\gamma_1\gamma_2}\cos(\Delta\theta_{12}),
\end{equation}
where $\Delta\theta_{12}=\theta_1-\theta_2$.

The corresponding Hamiltonian reads as:
\begin{equation}\label{eq:H2}
	\vb{H}=
	\begin{pmatrix}
		\Omega_1  &  -\iu\gamma_{12} \\
		  -\iu\gamma_{12}& \Omega_2
	\end{pmatrix},
\end{equation}
and its non-diagonal form determines that, according to Eq.~\eqref{eq:CMT1main}, even in the absence of external excitation, neither of the modes $\psi_1$ or $\psi_2$ evolves without mixing with the other. Therefore, they are not true eigenmodes in the classical sense, they are not those obtained by numerical eigenstate solvers or observed by the S-matrix poles in transmission and reflection. This is rarely discussed in the literature, but the fact itself is known for a long time, see e.g. \cite{Suh2004}.

To express the true modes, one resolves the Hamiltonian eigenfrequencies as:  \begin{equation}\label{eq:Omega_pm}
    \Omega_\pm=\frac{1}{2}(\Omega_1+\Omega_2)\pm\sqrt{\frac{1}{4}(\Omega_1-\Omega_2)^2-\gamma_{12}^2},
\end{equation}
and introduces the transformation matrix
\begin{equation}\label{eq:Tmatr}
    \vb{T}=\frac{1}{D}
    	\begin{pmatrix}
		-\iu(\Omega_2-\Omega_+)  &  \gamma_{12} \\
		  \gamma_{12} & \iu(\Omega_2-\Omega_+)
	\end{pmatrix}
\end{equation}
with
\begin{equation}\label{eq:D}
    D^2=(\Omega_+-\Omega_-)(\Omega_2-\Omega_+).
\end{equation}
which diagonalizes the Hamiltonian as
\begin{equation}
    \vb{T}^{\mathsf{T}}\vb{H}\vb{T}=\tilde{\vb{H}},
\end{equation}
with
\begin{equation}\label{eq:HD}
    \tilde{\vb{H}}=
    	\begin{pmatrix}
		\Omega_+ &  0\\
		  0 & \Omega_-
	\end{pmatrix}.
\end{equation}
This transformation matrix is symmetric and orthogonal, so that $\vb{T}^{\mathsf{T}}\vb{T}=\mathbf{I}$, but it is not Hermitian which reflects the non-Hermitian character of the Hamiltonian.

Accordingly, one can transform the CMT equations \eqref{eq:CMT1main} and \eqref{eq:CMT2main} to a new form with the diagonal Hamiltonian:
\begin{equation}\label{eq:CMT1diag}
	\frac{\dd\tilde{\boldsymbol{\psi}}}{\dd t } = \iu(\omega\mathbf{I}-\tilde{\vb{H}}) \tilde{\boldsymbol{\psi}}+\tilde{\vb{M}}^{\mathsf{T}}\vb{a},
\end{equation}
and
\begin{equation}\label{eq:CMT2diag}
	\vb{b} = \tilde{\vb{M}}\tilde{\boldsymbol{\psi}} + \vb{C}\vb{a}
\end{equation}
with the vector of new mode amplitudes introduced as:
\begin{equation}\label{eq:tildep}
	\tilde{\boldsymbol{\psi}} = \vb{T}\boldsymbol{\psi}.
\end{equation}
and the new matrix of coupling  parameters:
\begin{equation}\label{eq:tildeM}
	\tilde{\vb{M}} = \vb{M}\vb{T}.
\end{equation}
This transformed set of CMT equations allows expressing the S-matrix in the form of Eq.~\eqref{eq:Sorthog} explicitly describing the contributions of two resonances:
\begin{equation}\label{eq:SmatDiag}
	\vb{S}=\vb{C}-\frac{\vb{m}_+ \vb{m}_+^{\mathsf{T}}}{\iu(\omega-\Omega_+)}-\frac{\vb{m}_- \vb{m}_-^{\mathsf{T}}}{\iu(\omega-\Omega_-)},
\end{equation}
where the vectors of coupling parameters of the mixed modes constitute the coupling matrix:
\begin{equation}\label{eq:tildeMmatrix}
	\tilde{\vb{M}}=\left(\vb{m}_+,\vb{m}_-\right) =
	\begin{pmatrix}
		\tilde{m}_{+\mathrm{R}} & \tilde{m}_{-\mathrm{R}}\\
		\tilde{m}_{+\mathrm{L}} & \tilde{m}_{-\mathrm{L}}\\
		\tilde{m}_{+\mathrm{L}} & \tilde{m}_{-\mathrm{L}}\\
		\tilde{m}_{+\mathrm{R}} & \tilde{m}_{-\mathrm{R}}\\
	\end{pmatrix},
\end{equation}
where
\begin{align}
    \tilde{m}_{+\mathrm{R},\mathrm{L}}&=\frac{1}{D}\left[-\iu(\Omega_2-\Omega_+)m_{1\mathrm{R},\mathrm{L}}+\gamma_{12}m_{2\mathrm{R},\mathrm{L}}\right],\label{eq:tildem+}\\
    \tilde{m}_{-\mathrm{R},\mathrm{L}}&=\frac{1}{D}\left[\gamma_{12}m_{1\mathrm{R},\mathrm{L}}+\iu(\Omega_2-\Omega_+)m_{2\mathrm{R},\mathrm{L}}\right], \label{eq:tildem-}
\end{align}
where the parity-determined relations \eqref{eq:meven} imposed the same restriction on the mixed modes: a mixture of even modes is also even.
Non-Hermitian mode transformation $\vb{T}$ does not preserve the mode energy. Accordingly, the relations \eqref{eq:timereversalconstraint} and \eqref{eq:energyconstraint} do not restrict the coupling parameters \eqref{eq:tildem+} and \eqref{eq:tildem-} and do not dictate the mixed modes to be linearly polarized in the far-field. Elliptically and, eventually, circularly polarized modes become possible.

Expressing the S-matrix in the form \eqref{eq:SmatDiag} allows, in particular, writing its components describing the cross-polarized transmission as:
\begin{align}
    t_{\text{RL}}&=-\frac{\tilde{m}_{+\mathrm{L}}^2}{\iu(\omega-\Omega_+)}-\frac{\tilde{m}_{-\mathrm{L}}^2}{\iu(\omega-\Omega_-)},\label{eq:tRLtLR}\\
    t_{\text{LR}}&=-\frac{\tilde{m}_{+\mathrm{R}}^2}{\iu(\omega-\Omega_+)}-\frac{\tilde{m}_{-\mathrm{R}}^2}{\iu(\omega-\Omega_-)}.
\end{align}
To characterize a particular dressed mode, one can take the ellipticity of its far-field, the Stokes parameter $S_3$:
\begin{equation}
    S_3^\mathrm{\pm mode} = \frac{|\tilde{m}_{\pm \mathrm{R}}|^2 - |\tilde{m}_{\pm \mathrm{L}}|^2}{|\tilde{m}_{\pm \mathrm{R}}|^2 + |\tilde{m}_{\pm \mathrm{L}}|^2}.
\label{eq:S3mode}
\end{equation}
as an indicator of the mode chirality.
It reaches its extreme values $\pm 1$ when either of the coupling parameters vanishes.

It is truly remarkable that the conversion circular dichroism remains zero in the absence of dissipation loss even if one of the dressed modes eventually becomes circularly polarized. This general fact directly follows from the S-matrix form~\eqref{eq:SmatDiag}, which has no contributions from the background to the coefficients of reflections  $r_{\text{RR}}$ and $r_{\text{LL}}$ and transmissions $t_{\text{LR}}$ and $t_{\text{RL}}$. As the dressed modes ``$\pm$'' inherit the same parity from the bare modes, the magnitudes of these coefficients are equal in pairs: $|t_{\text{LR}}|=|r_{\text{RR}}|$ and $|t_{\text{RL}}|=|r_{\text{LL}}|$. The energy conservation dictates that
\begin{align}
    1&=T_{\mathrm{LR}}+T_{\mathrm{RR}}+R_{\mathrm{RR}}+R_{\mathrm{LR}},\\
    1&=T_{\mathrm{RL}}+T_{\mathrm{LL}}+R_{\mathrm{LL}}+R_{\mathrm{RL}},
\end{align}
where $r_{\text{LR}}=r_{\text{RL}}$ due to reciprocity and $t_{\text{RR}}=t_{\text{LL}}$ due to the membrane plane symmetry. Accordingly, the cross-polarized transmissions are bound to be equally strong, $T_{\mathrm{RL}}=T_{\mathrm{LR}}$.

\section{Role of absorption losses}
\label{app:losses}

Section~\ref{sec:losses} gives the main results. Here we derive them and show the full-wave spectra. As in Sec.~\ref{sec:losses}, we add small absorption through the substitution $\vb{\Omega}-\iu\vb{\Gamma}\to\vb{\Omega}-\iu(\vb{\Gamma}+\vb{\Gamma}_{\mathrm{abs}})$, where $\vb{\Gamma}_{\mathrm{abs}}=\operatorname{diag}(\gamma_1^{\mathrm{abs}},\gamma_2^{\mathrm{abs}},\dots)$. The coupling matrix $\vb{M}$ does not change in the fist approximation. Thus $\gamma_n$ remains the radiative decay rate of mode $n$, and its total decay rate is $\gamma_n^{\mathrm{tot}}=\gamma_n+\gamma_n^{\mathrm{abs}}$.

\subsection{Port-orthogonal mechanism}
\begin{figure*}
    \centering
    \includegraphics[width=0.8\linewidth]{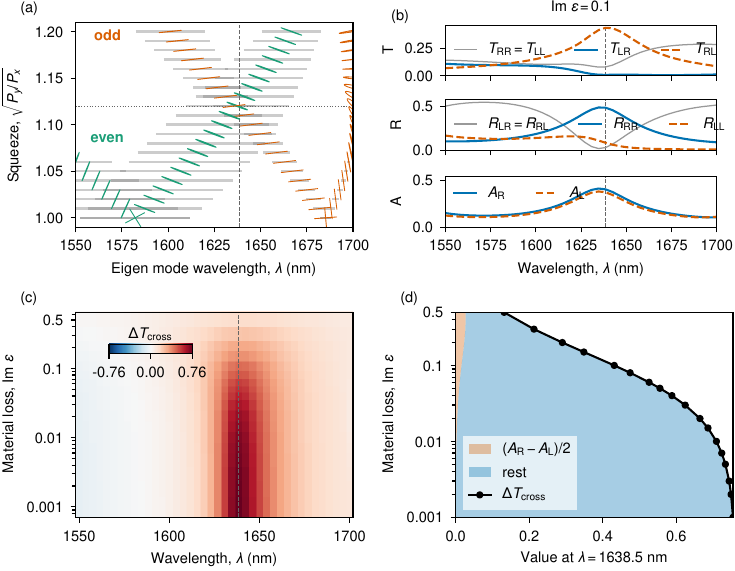}
    \caption{
    \textbf{Losses in the port-orthogonal mechanism}, full-wave results for the crossing of Fig.~\ref{fig:squeeze} (same parameters).
    (a)~Eigenmodes as in Fig.~\ref{fig:losses}(a), all sweep values. The dotted line marks $\sqrt{P_y/P_x}=1.12$, where we calculate the spectra. The dashed line marks the wavelength $\lambda\approx1635\nm$ of the largest $|\Delta T_{\text{cross}}|$.
    (b)~Transmittance $T$, reflectance $R$, and absorptance $A$ for RCP (solid) and LCP (dashed) incidence at $\Im\varepsilon=0.1$. The first index is the output helicity, and the second index is the input helicity.
    (c)~$\Delta T_{\text{cross}}$ versus the wavelength and the material loss $\Im\varepsilon$.
    (d)~$\Delta T_{\text{cross}}$ at the fixed wavelength $\lambda\approx1635\nm$ (dashed line) versus $\Im\varepsilon$ (points), split into $(A_{\mathrm{R}}-A_{\mathrm{L}})/2$ (orange) and the rest (blue).
    }
    \label{fig:losses_orthog}
\end{figure*}

For opposite-parity modes, the matrix $\vb{\Omega}-\iu\vb{\Gamma}-\iu\vb{\Gamma}_{\mathrm{abs}}$ stays diagonal. Equation~\eqref{eq:Sorthog_main} still applies, with $\gamma_n\to\gamma_n^{\mathrm{tot}}$ in the denominators only. The amplitudes~\eqref{eq:tLR_orthog} become
\begin{equation}
    B_n=\frac{\eta_n}{2}\,\eu^{\iu(\alpha+p_n\beta)}L_n(\omega),
    \quad
    \eta_n=\frac{\gamma_n}{\gamma_n^{\mathrm{tot}}},
    \quad
    L_n(\omega)=\frac{\gamma_n^{\mathrm{tot}}}{\iu(\omega-\omega_n)-\gamma_n^{\mathrm{tot}}}.
    \label{eq:An_loss}
\end{equation}
Here $\eta_n\leq 1$ is the radiative efficiency of mode $n$, the fraction of its decay that goes into the ports. The function $L_n$ is a Lorentzian with unit peak, $L_n(\omega_n)=-1$. With $p_2=-p_1$, Eq.~\eqref{eq:dT_orthog} gives
\begin{equation}
    \Delta T_{\mathrm{cross}}=-\eta_1\eta_2\,\sin(2\Delta\theta_{12})\,\Im\!\left[\eu^{2\iu p_1\beta}L_1L_2^*\right].
    \label{eq:dT_orthog_loss}
\end{equation}
Absorption therefore has two effects. It multiplies $\Delta T_{\mathrm{cross}}$ by $\eta_1\eta_2$, and it broadens both resonances. Because $|L_n|\leq 1$, the bound $|\Delta T_{\mathrm{cross}}|\leq 1$ becomes $|\Delta T_{\mathrm{cross}}|\leq\eta_1\eta_2$. For degenerate modes with $\omega_1=\omega_2=\omega_0$ and $\gamma_1^{\mathrm{tot}}=\gamma_2^{\mathrm{tot}}=\gamma^{\mathrm{tot}}$,
\begin{equation}
    \Delta T_{\mathrm{cross}}=-\operatorname{sign}(p_1)\,\eta_1\eta_2\,\frac{(\gamma^{\mathrm{tot}})^2}{(\gamma^{\mathrm{tot}})^2+(\omega-\omega_0)^2}\,\sin(2\beta)\sin(2\Delta\theta_{12}).
    \label{eq:dT_degenerate_loss}
\end{equation}
For $\eta_1=\eta_2=1$, this is Eq.~\eqref{eq:dT_degenerate}. On resonance of degenerate modes, $\omega=\omega_1=\omega_2$, we have $L_1=L_2=-1$ for any $\gamma_n^{\mathrm{tot}}$, and Eq.~\eqref{eq:dT_orthog_loss} reduces to Eq.~\eqref{eq:loss_orthog_main}. The port-orthogonal mechanism does not need absorption, and a lossless membrane gives the largest signal. Figure~\ref{fig:losses_orthog} shows the full-wave spectra for the crossing of Fig.~\ref{fig:squeeze}. At the resonance wavelength, $\Delta T_{\text{cross}}$ decreases monotonically with the loss, and $(A_{\mathrm{R}}-A_{\mathrm{L}})/2$ stays negligible.

\subsection{Port-non-orthogonal mechanism}
Now all modes in the spectral range have the same parity $p$. Equation~\eqref{eq:parity_main} gives $m'_{n\mathrm{L}}=p\,m_{n\mathrm{R}}$ and $m'_{n\mathrm{R}}=p\,m_{n\mathrm{L}}$ for every mode $n$. Thus the row of $\vb{M}$ for the port $b'_{\mathrm{L}}$ is $p$ times the row for $b_{\mathrm{R}}$. Likewise, the row for $b'_{\mathrm{R}}$ is $p$ times the row for $b_{\mathrm{L}}$. The resonant term of the S-matrix~\eqref{eq:SmatCMTmain} keeps this relation for any mode Hamiltonian, with or without $\vb{\Gamma}_{\mathrm{abs}}$. The isotropic background~\eqref{eq:Cmain} does not contribute to $r_{\mathrm{RR}}$, $t_{\mathrm{LR}}$, $r_{\mathrm{LL}}$, and $t_{\mathrm{RL}}$. Hence
\begin{equation}
    t_{\mathrm{LR}}=p\,r_{\mathrm{RR}},
    \qquad
    t_{\mathrm{RL}}=p\,r_{\mathrm{LL}}.
    \label{eq:tr_sameparity}
\end{equation}
Let $A_{\mathrm{R}}$ ($A_{\mathrm{L}}$) be the absorbed fraction of the incident RCP (LCP) power. Energy conservation for each incident helicity reads
\begin{equation}
\begin{split}
    1&=T_{\mathrm{LR}}+T_{\mathrm{RR}}+R_{\mathrm{RR}}+R_{\mathrm{LR}}+A_{\mathrm{R}},\\
    1&=T_{\mathrm{RL}}+T_{\mathrm{LL}}+R_{\mathrm{LL}}+R_{\mathrm{RL}}+A_{\mathrm{L}}.
\end{split}
\end{equation}
The up-down mirror symmetry gives $T_{\mathrm{RR}}=T_{\mathrm{LL}}$, and reciprocity gives $R_{\mathrm{LR}}=R_{\mathrm{RL}}$. We subtract the two equations and use $T_{\mathrm{LR}}=R_{\mathrm{RR}}$ and $T_{\mathrm{RL}}=R_{\mathrm{LL}}$ from Eq.~\eqref{eq:tr_sameparity}. This gives
\begin{equation}
    \Delta T_{\mathrm{cross}}=\frac{A_{\mathrm{R}}-A_{\mathrm{L}}}{2}.
    \label{eq:dT_abs}
\end{equation}
For a lossless membrane, $A_{\mathrm{R}}=A_{\mathrm{L}}=0$, and Eq.~\eqref{eq:dT_abs} gives the result $\Delta T_{\mathrm{cross}}=0$ of Section~\ref{app:SmatPortNonOrthog}. With absorption, the conversion dichroism is equal to one half of the circular dichroism in absorption. The difference $A_{\mathrm{R}}-A_{\mathrm{L}}$ needs a dressed mode that couples differently to RCP and LCP waves, that is, an elliptically polarized dressed mode. %
The up-down mirror symmetry limits $A_{\mathrm{R,L}}\leq 1/2$, which gives the bound $|\Delta T_{\mathrm{cross}}|\leq 1/4$~\eqref{eq:dT_abs_bound_short} (Sec.~\ref{sec:losses}). The bound needs critical coupling, so $\Delta T_{\mathrm{cross}}$ has its maximum at a finite loss.

\begin{figure*}
    \centering
    \includegraphics[width=0.8\linewidth]{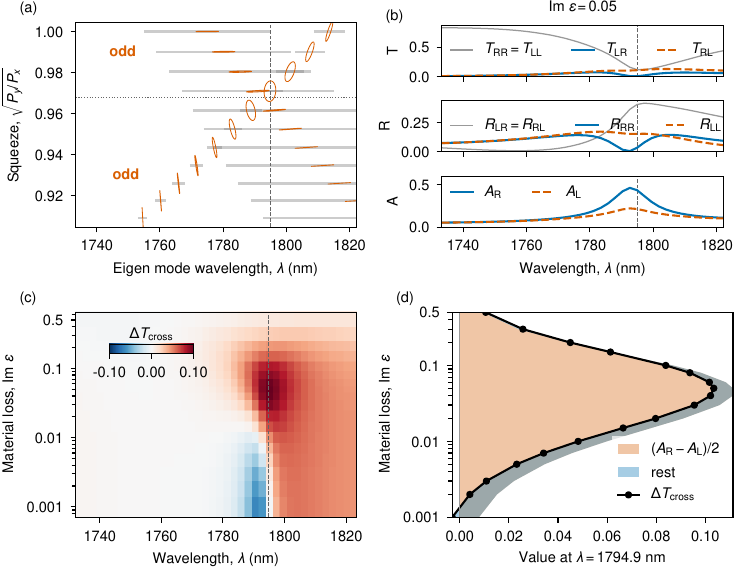}
    \caption{
    \textbf{Losses in the port-non-orthogonal mechanism}, full-wave results for the same-parity crossing of Fig.~\ref{fig:losses}(b) (same parameters).
    (a)~Eigenmodes as in Fig.~\ref{fig:losses}(b). The dotted line marks $\sqrt{P_y/P_x}=0.968$, where we calculate the spectra. The dashed line marks the wavelength $\lambda=1794.9\nm$ of the largest $|\Delta T_{\text{cross}}|$.
    (b)~Transmittance $T$, reflectance $R$, and absorptance $A$ for RCP (solid) and LCP (dashed) incidence at $\Im\varepsilon=0.05$, the loss of the largest $|\Delta T_{\text{cross}}|$. Indices as in Fig.~\ref{fig:losses_orthog}.
    (c)~$\Delta T_{\text{cross}}$ versus the wavelength and the material loss $\Im\varepsilon$.
    (d)~$\Delta T_{\text{cross}}$ at the fixed wavelength $\lambda=1794.9\nm$ (dashed line) versus $\Im\varepsilon$ (points), split into $(A_{\mathrm{R}}-A_{\mathrm{L}})/2$ (orange) and the rest (blue). Gray marks the overlap where the rest is negative.
    }
    \label{fig:losses_nonorthog}
\end{figure*}

Figure~\ref{fig:losses_nonorthog} shows the full-wave spectra for the same-parity crossing of Fig.~\ref{fig:losses}(b). Without loss, $\Delta T_{\text{cross}}$ is small and comes from an even mode outside the spectral window, through the port-orthogonal mechanism. With loss, $\Delta T_{\text{cross}}$ at the crossing becomes positive, and $(A_{\mathrm{R}}-A_{\mathrm{L}})/2$ accounts for almost all of it. For strong loss, $A_{\mathrm{R}}$ saturates close to the limit $1/2$, while $A_{\mathrm{L}}$ continues to grow. Thus $\Delta T_{\text{cross}}$ has its maximum at a finite loss and then decreases.

\section{Numerical methods}

Eigenmodes and spectra are obtained with the finite-element solver COMSOL Multiphysics (Wave Optics module, \texttt{ewfd}). A single unit cell with Floquet-periodic boundaries along $x$ and $y$ and perfectly matched layers along $z$ represents the infinite membrane. An eigenfrequency study yields the complex eigenfrequencies $f_n$. We use the angular frequencies $\Omega_n=2\pi f_n=\omega_n-\iu\gamma_n$ (conjugated to the $\eu^{-\iu\omega t}$ convention), given in units of $10^{12}$\,rad/s throughout and the modal fields $\vb{E}_n$.

The transmission amplitudes of the S-matrix~\eqref{eq:Smatrix} are computed with a background-field formulation. The background field is set to a circularly polarized (RCP or LCP) plane wave incident along $-z$, and the solver returns the resulting total field. The co- and cross-polarized transmission coefficients are obtained by projecting the field onto the circular basis~\eqref{eq:circbasis} and averaging it over a transverse plane on the transmission side, placed a small gap before the PML so that only the propagating zeroth diffraction order contributes; normalizing to the incident amplitude gives $t_{\mathrm{RR}}, t_{\mathrm{LL}}, t_{\mathrm{RL}}, t_{\mathrm{LR}}$ and hence $\Delta T_{\mathrm{cross}}$~\eqref{eq:dTcross}.

Eqs.~\eqref{even}--\eqref{odd} for the up--down parity $p_n$ can be compactly rewritten using the $z\to-z$ operator $\hat{\sigma}_z = \operatorname{diag}(1, 1, -1)$ as~\cite{Kumar2025ACSPhot}
\begin{equation}
    \hat{\sigma}_z \vb{E}_n (\hat{\sigma}_z^{-1} \vb{r}) = p_n \vb{E}_n (\vb{r}).
\end{equation}
We can convert this to the normalized overlap of a mode with its mirror image for the numerical calculation:
\begin{equation}
    p_n = \frac{\int_V \hat{\sigma}_z \vb{E}_n (\hat{\sigma}_z^{-1} \vb{r}) \cdot \vb{E}_n(\vb{r}) \dd V}{\int_V \vb{E}_n (\vb{r}) \cdot \vb{E}_n(\vb{r}) \dd V}.
\end{equation}
The integral runs over the membrane domain $V$, taken symmetric about $z=0$. In COMSOL the mirrored field $\vb{E}_n(\hat{\sigma}_z^{-1}\vb{r})$ is evaluated with a General Extrusion operator implementing the map $(x,y,z)\to(x,y,-z)$, and the overlap with an integration coupling over $V$. The normal component $E_{n,z}$ enters with a minus sign, as dictated by $\hat{\sigma}_z$.

\end{document}